\documentclass[aps,prl,10pt,twocolumn,superscriptaddress,longbibliography,nofootinbib,floatfix]{revtex4-2}
\usepackage{graphicx}
\usepackage{amsmath,amssymb}
\usepackage[dvipsnames]{xcolor}
\usepackage[colorlinks, linkcolor=BrickRed,anchorcolor=blue,citecolor=blue,urlcolor=blue]{hyperref}
\usepackage{xurl}
\usepackage[capitalize]{cleveref}

\begin{document}

\title{Evaluating Verified Autonomy in Quantum Engineering}
\author{Naixu Guo$^{*,\dagger}$}
\affiliation{Centre for Quantum Technologies, National University of Singapore, 117543, Singapore}
\author{Changhao Li$^{*,\dagger}$}
\affiliation{Unitary Foundation, San Francisco, CA, USA}
\author{Siyu Cheng$^{*}$}
\affiliation{Department of Physics, Boston College, Chestnut Hill, Massachusetts 02467, USA}
\author{Qicheng Tang$^{*}$}
\affiliation{School of Physics, Georgia Institute of Technology, Atlanta, GA 30332, USA}
\author{Binzhao Luo}
\affiliation{GaugeForge PTE. LTD., 049422, Singapore}
\author{Bikun Li}
\affiliation{Chicago Quantum Institute and Pritzker School of Molecular Engineering, University of Chicago, Chicago, Illinois 60637, USA}
\author{Yuxuan Du}
\affiliation{College of Computing and Data Science, Nanyang Technological University, Singapore 639798, Singapore}
\author{Shihao Ru$^{\dagger}$}
\affiliation{School of Electrical and Electronic Engineering, Nanyang Technological University, Singapore 639798, Singapore}
\author{Jiaqi Cai$^{\dagger}$}
\affiliation{GaugeForge PTE. LTD., 049422, Singapore}

\date{\today}

\begin{abstract}
Reliable quantum engineering is essential for turning quantum phenomena into practical technologies.
As quantum platforms grow in scale and complexity, their characterization and operation require increasing human effort and coordination.
Scientific artificial intelligence agents, which can plan experiments, operate instruments, and analyze observations, offer a promising route towards autonomous quantum engineering. 
Yet whether current agents can perform reliably in this setting has not been systematically established.
To fill this gap, we developed Quantum-Harbor, a virtual laboratory that provides a controlled execution environment for agents to interact with quantum systems. 
This design enables direct verification of both the actions taken and the conclusions drawn.
Building on this framework, we introduce QIQCBench, a benchmark of $49$ expert-authored tasks spanning multiple layers including calibration and control, error correction and compilation, sensing and networking.
Across $17$ frontier agentic systems, QIQCBench reveals wide variation in verified performance.
These results expose a substantial gap between demonstrating capability and achieving reliable operation, and establish Quantum-Harbor as a foundation for measuring progress towards verified autonomy in quantum engineering.

\end{abstract}
\maketitle

{\let\thefootnote\relax\footnotetext{$^{*}$~These authors contributed equally to this work.}}
{\let\thefootnote\relax\footnotetext{$^{\dagger}$~Corresponding authors: naixug@u.nus.edu (N.G.), changhaoli96@gmail.com (C.L.), shihao.ru.2018@gmail.com (S.R.), jiaqi.cai@gauge-forge.com (J.C.)}}

\section{Introduction}\label{sec:intro}
\begin{figure*}[htbp]
\centering
\includegraphics[width=\textwidth]{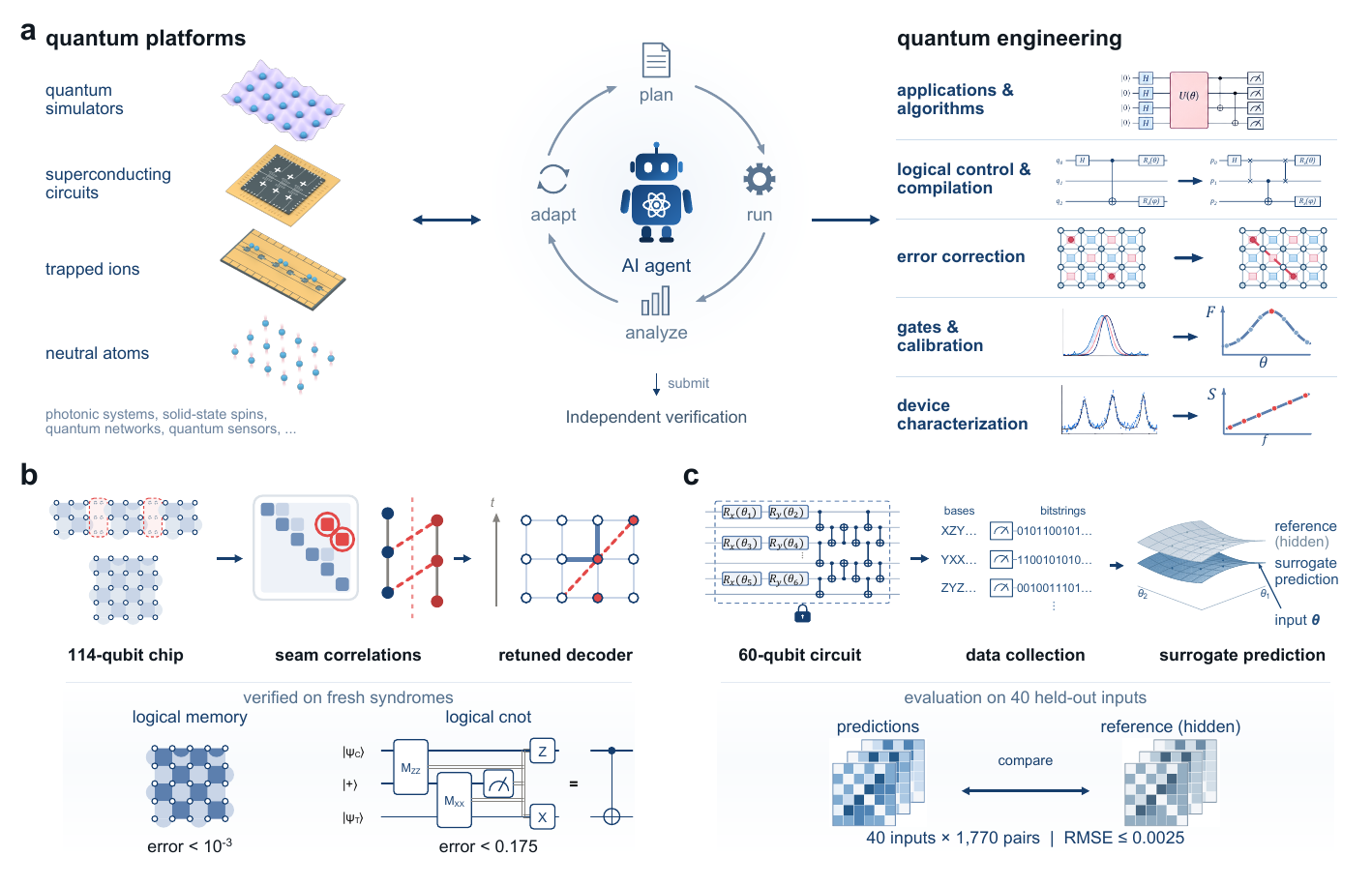}
\caption{AI agents for quantum engineering and representative workflows.
\textbf{a}.~Agent-operated quantum engineering. The agent iterates planning, execution, analysis, and adaptation while interacting with quantum laboratories. The workflow operates across the layers shown on the right, with individual tasks typically engaging multiple layers, and its outputs are submitted for independent verification.
\textbf{b}.~Decoder calibration for a lattice-surgery logical CNOT gate. 
The agent measures error correlations across the merge seam between surface-code patches and submits a retuned decoder, verified on fresh syndrome records.
The public acceptance criteria require a distance-5 logical-memory error below \(10^{-3}\) per cycle and, for eight logical-CNOT test circuits spanning both measurement bases, a mean error below \(0.175\) with each circuit-specific error below \(0.185\).
\textbf{c}.~Surrogate modeling of a hidden 60-qubit parameterized quantum circuit.
The agent collects measurement data and constructs a surrogate model.
Once the measurement record is fixed, 40 held-out inputs are revealed, for which the agent predicts 1,770 two-qubit correlations for each input.
The predictions are compared with hidden reference correlations, with accuracy evaluated by the root mean square error.
}
\label{fig:overview}
\end{figure*}
Quantum technologies, spanning computing, simulation, communication, and sensing~\cite{arute2019quantum,daley2022practical, koji2023repeaters,degen2017sensing}, are moving from proof-of-principle demonstrations toward practical applications~\cite{acharya2024threshold,bluvstein2024logical, paetznick2026logical,kim2023utility,google2025echoes, shaw2024benchmarking,liu2026longlived, zheng2026largescale, stas2026entanglement,zhou2025sensing, rovny2025multi,Ji2024NaturePhotonics}.
Quantum engineering enables this transition by integrating quantum platforms and control methods as well as software into reliable systems.
Building and operating quantum platforms has traditionally relied heavily on human experts, who assess device information, interpret experimental results, and make the necessary adjustments~\cite{alexeev2025artificial}. 
As these platforms scale, the number of components, parameters, and workflows that must be coordinated and optimized grows, making manual diagnosis increasingly impractical~\cite{beverland2022assessing,mohseni2024supercomputer}.
Several individual engineering tasks have therefore been automated~\cite{moon2020machine, chong2017programming, wang2017hamiltonian}.
Scientific artificial intelligence (AI) agents offer a way to extend this automation to longer workflows~\cite{alexeev2025artificial, du2026artificial}.
These agents can plan actions, use experimental and computational tools, and adapt their strategies in response to feedback~\cite{yao2023react}.
Recent efforts have begun to apply such agents to quantum experiments and software workflows~\cite{cao2025kagents,xu2026vibecal,arlt2025agents,isogawa2026sensing,dalyac2026lowering,fu2025qagent,li2026autosim, gustin2026agente, shiraishi2026model}.
Yet whether current agents can reliably carry out complete quantum engineering workflows remains unclear.

Evaluating this capability requires more than checking whether an agent produces the correct answer or completes a predefined task.
In practical quantum engineering, critical information often becomes available only during operation~\cite{proctor2020drift, sivak2026reinforcement}.
Because device information may be only partially known, an agent must choose and execute experiments or computations, interpret the resulting observations, and decide how to proceed~\cite{lennon2019efficiently, baum2021experimental}. 
Even with the relevant scientific knowledge, an agent may rely on information it did not measure, misinterpret experimental data, or produce a result that fails on fresh data~\cite{cao2026qcaleval}.
We therefore focus on \emph{verified autonomy}, the ability of an agent to complete an engineering task and support its conclusions with evidence generated during execution.
Outputs that cannot be verified directly from recorded data, such as calibrations or designs, must instead be tested under fresh hidden conditions.
Existing benchmarks typically assess final outputs without systematically checking whether those results are supported by the agent's execution record~\cite{minami2025quantumbench, yang2024qcircuitbench}.

To make verified autonomy measurable, we develop Quantum-Harbor, a virtual quantum laboratory that provides experimental and computational interfaces while keeping device properties and grading ground truth hidden.
For each task, the agent receives only a high-level objective and a finite action budget.
The agent must decide how to interact with the environment, conduct experiments or computations, and submit a final result. 
Quantum-Harbor records the agent's actions and the resulting data and uses this record to verify the submitted result. 
The grader verifies numerical results by recomputing them from raw experimental data, while for calibration and design outputs, it retests them under fresh hidden conditions.
Building on this environment, we construct a benchmark comprising 49 expert-designed quantum engineering tasks covering hardware and software workflows at multiple architectural layers.
We use this benchmark to systematically evaluate verified autonomy in frontier agentic AI systems.

\section{Results}\label{sec:results}
\subsection{Representative quantum research tasks}

Agent-operated quantum engineering places the AI agent in direct interaction with quantum laboratories, closing the loop between experiments and analysis (Fig.~\ref{fig:overview}a).
The laboratory may host quantum simulators or quantum platforms such as superconducting circuits, trapped ions, neutral atoms and photonic systems.
During execution, the agent chooses experimental or computational actions, interprets the resulting data, and revises its plan as new information is acquired.
The workflow therefore evolves adaptively as execution progresses.
A single workflow may span multiple layers of quantum engineering, from device characterization and gate calibration to error correction, compilation, and applications.
The resulting execution record provides the evidence needed to assess and verify the agent's outputs.
We now examine two representative quantum research tasks that require distinct agent capabilities (Fig.~\ref{fig:overview}b,c).

The first task focuses on the logical CNOT gate which is a basic building block of fault-tolerant quantum computation. Lattice surgery realizes this gate by merging and splitting error-correcting code patches~\cite{horsman2012,litinski2019}.
Recent experiments have demonstrated this approach in trapped-ion~\cite{erhard2021} and superconducting processors~\cite{besedin2026,lin2026surgery}.
However, tuning the decoder to account for correlated noise not captured by hardware datasheets remains an expert-driven process~\cite{acharya2024threshold,bausch2024alphaqubit}.
In this task, the agent operates a simulated 114-qubit transmon grid containing three distance-3 surface-code patches, a routing region used to fuse them during surgery, and a distance-5 memory patch. It submits decoder configurations that the hidden grader replays on fresh syndrome data.
A successful configuration must clear two criteria that bound the per-cycle error of the distance-5 memory and the logical-CNOT error across test circuits in both measurement bases, as shown in Fig.~\ref{fig:overview}b.
Both criteria follow from a hidden noise model that is set at the physical error rates of current below-threshold processors~\cite{acharya2024threshold}.
The task also embeds a failure mode taken from real laboratories.
The lab notebook records earlier memory measurements on separate patches and recommends reusing the resulting decoder calibration for the logical operation.
Merging the patches introduces correlated errors along their shared boundary, which are not visible in those earlier measurements, so the calibrated decoder can fail during surgery.
The strongest agents instead measure the merged configuration directly and clear both criteria, with the best run reaching a memory error of $6.65\times10^{-4}$ per cycle and a mean logical-CNOT error of $0.131$ on fresh syndrome records.

The surrogate model task~\cite{schreiber2023classical, du2025efficient, liao2026demonstration} requires the agent to design informative measurement protocols and construct a predictive model of an unknown quantum system (Fig.~\ref{fig:overview}c). 
The agent interacts with a 60-qubit circuit $U(\boldsymbol{\theta})$ containing Clifford and non-Clifford gates, with 18 non-Clifford rotation gates controlled by six independent parameters.
Within a fixed experimental budget, it determines how to sample the parameter space and allocate shots across measurement bases. 
After the measurement phase ends, 40 previously hidden inputs are revealed, for which the agent must predict $1{,}770$ pairwise correlations without further measurements. 
The predictions are compared with reference values generated from the same circuit and readout noise model.
Success therefore requires learning from limited measurements and generalizing to unseen parameter settings.
Failed runs often expend substantial experimental resources but fall short because of uninformative sampling, unreliable validation or incomplete execution. 
Many of these runs progress through much of the workflow, while consistent end-to-end completion remains challenging.

These examples illustrate both the emerging capabilities and current limitations of agentic quantum engineering. The lattice surgery task requires identifying noise that emerges only in the merged configuration, whereas the surrogate model task requires learning from limited measurements to predict unseen inputs. In both cases, success depends on information acquired during execution. Evaluating such workflows therefore requires preserving the interaction record and, where conclusions cannot be checked directly, testing them on fresh data or unseen inputs. These requirements motivate a controlled environment for systematic evaluation.
\subsection{The Quantum-Harbor framework}

\begin{figure*}[t]
\centering
\includegraphics[width=0.9\textwidth]{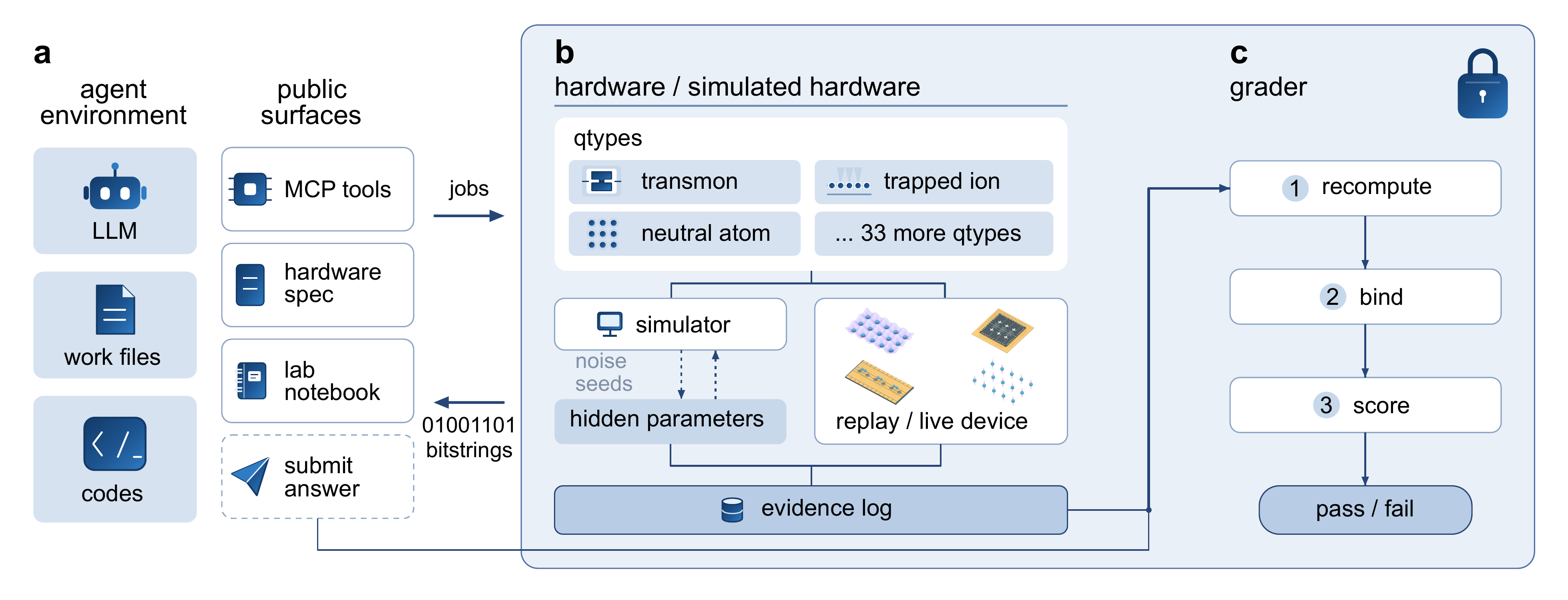}
\caption{Architecture and evidence-bound grading in Quantum-Harbor.
  \textbf{a.}~The agent environment contains the language model, work
  files, and agent-written code. Through the public interfaces, the
  agent consults hardware specifications and lab notebooks, submits
  experiment jobs, retrieves raw measurement records (bitstrings shown),
  and submits its final answer. Experiment jobs execute asynchronously.
  \textbf{b.}~The hardware layer provides quantum hardware abstractions
  (\texttt{qtypes}) for transmons, neutral atoms, trapped ions, and other
  platforms. Experiments run on simulators governed by hidden device
  and noise parameters, or on replay and live-device backends where
  supported. An evidence log records experimental requests and outcomes
  for subsequent verification.
  \textbf{c.}~The grader recomputes relevant quantities, binds submitted
  claims to the recorded experiments, and applies task-specific scoring
  criteria (pass/fail shown). For calibration tasks, submitted
  configurations can be evaluated on fresh simulated data.
  The shaded region and dashed boundary indicate components inaccessible
  to the agent, including hidden parameters, the private evidence log,
  and grading logic.}
\label{fig:framework}
\end{figure*}

Quantum-Harbor provides the large language model (LLM) agent with a sandbox environment to interact with quantum hardware and complete quantum research tasks (Fig.~\ref{fig:framework}). It builds on Harbor~\cite{harbor2026}, a general framework for sandboxed agent tasks such as terminal shell use and software development~\cite{merrill2026terminalbench,terminalbenchscience2026}. Quantum-Harbor adds three features to Harbor: a laboratory interface for quantum experiments, interchangeable quantum hardware backends, and evidence-bound grading. Each Quantum-Harbor run launches two isolated environments, namely the agent environment and the sidecar. The agent environment (Fig.~\ref{fig:framework}a) runs the language model harness such as Codex and Claude Code, and stores the workfiles used by the agent. The agent reads the working files and writes code to interact with the public surface exposed in the agent environment.
The public surface is a set of tools exposed via the Model Context Protocol (MCP)~\cite{mcp2024}, a published standard that lets a language model call an external service the way a program calls a function. Through them the agent can read the hardware spec, write the lab notebook whose calibration entries may be outdated, and submit experiment jobs, and its final answer. The agent can also carry out the experiment through the Qiskit~\cite{javadiabhari2024quantum} or QCoDeS~\cite{qcodes} libraries that a human experimentalist would use. Every experiment submission returns a job handle to the agent and completes asynchronously, and the job results are raw records, such as readout points or per-shot bitstrings. 

Behind the public surface is the sidecar (Fig.~\ref{fig:framework}b,c). The agent can only access and interact with the public surface, and the sidecar is hidden and isolated from the agent. The sidecar hosts the detailed implementation of the quantum hardware or simulated hardware. To support a variety of quantum platforms, the sidecar contains different quantum hardware types called \texttt{qtypes} (Fig.~\ref{fig:framework}b) to accommodate the variety among different quantum hardware. A \texttt{qtype} abstracts the physics of one class of experimental platform, such as a multilevel transmon under pulse control, a trapped-ion chain, or a neutral-atom array. \texttt{Qtypes} can simulate quantum hardware, with predefined hidden parameters such as the coherence times, anharmonicities, readout confusion matrices, noise correlations, and random seeds (see Supplementary Material for details about the implementation of simulation). It also allows live connection to real quantum hardware, such as superconducting systems from IBM, Rigetti, and ion trap systems from IonQ and Quantinuum. The replay backend is the dry-run mode for the live quantum hardware and runs on recorded data. The simulator, real quantum hardware, and the replay backend are interchangeable. Switching among them within one \texttt{qtype} does not require any change to task definitions or public surfaces. 

The evidence log inside the sidecar records every experiment job executed on either simulator or real quantum hardware. An LLM might report a number it never measured~\cite{islam2026gwbenchmarks,chowdhury2025truthfulness}. Therefore, the grader (Fig.~\ref{fig:framework}c) is used to evaluate LLMs' performance on quantum research tasks and is bound to the evidence log. The grader verifies the task result by recomputing it from the raw data in the evidence log. When the submission is a calibration and not a number, the grader replays it against fresh draws from the hidden parameters. This prevents the agent from being rewarded for submitting false results. 
\subsection{The QIQCBench benchmark}

\begin{figure*}[t]
\centering
\includegraphics[width=\linewidth]{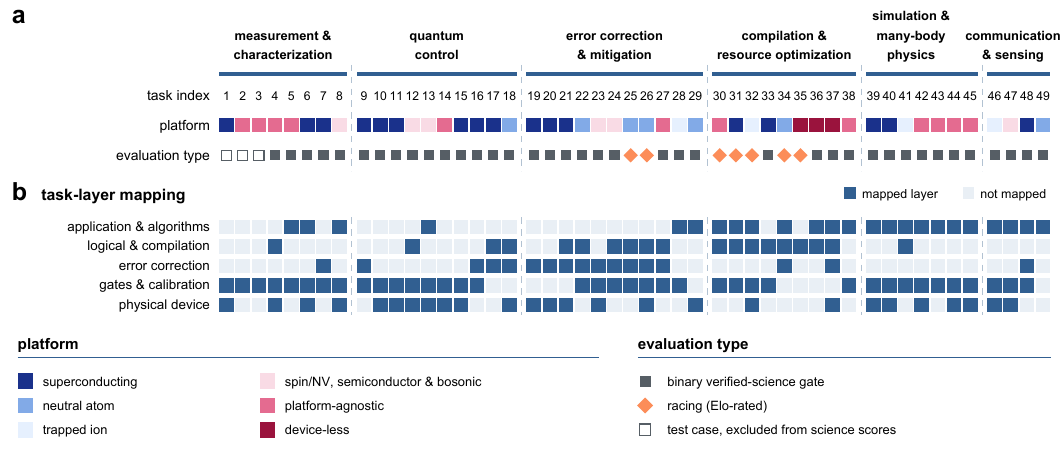}
\caption{Task composition and quantum computing stack coverage of QIQCBench.
\textbf{a.}~The 49 tasks grouped into six scientific categories. Each column represents one task. Colors identify the hardware platform, with separate labels for platform-agnostic and device-less tasks.
  Evaluation types are indicated by filled squares for the 39 tasks
  assessed by a binary pass--fail evaluation, orange diamonds for the
  7 racing tasks ranked by Elo rating, and open squares for the
  3 infrastructure test cases excluded from science scores.
\textbf{b.}~Mapping of each task to five layers of the quantum computing
  stack, adapted from Ref.~\cite{Jones2012layered}: physical device,
  gates and calibration, error correction, logical and compilation,
  and application and algorithms. Dark blue cells mark the layers
  engaged by each task. Pale cells indicate unmapped layers.
  Columns share the task indices and category boundaries of panel~\textbf{a},
  showing how individual tasks connect multiple layers of quantum engineering.}
\label{fig:distribution}
\end{figure*}

Building on the Quantum-Harbor framework, we next introduce QIQCBench, which comprises 49 tasks built on 36 qtypes, all contributed and reviewed by domain experts.
Their expertise spans the working breadth of quantum engineering. 
Fig.~\ref{fig:distribution}a shows the composition of the suite, with tasks organized into six categories.
Representative measurement and characterization tasks include estimating the relaxation time of a drifted transmon and reconstructing correlated gate errors~\cite{proctor2020drift,sarovar2020crosstalk}, while quantum control tasks range from pulse design to Hamiltonian engineering~\cite{koch2022quantumcontrol,baum2021experimental}.
Error correction and mitigation tasks include problems such as decoder calibration and logical-memory protection~\cite{acharya2024threshold,bausch2024alphaqubit}, and compilation and resource-optimization tasks map target computations to trapped-ion, neutral-atom, and superconducting architectures~\cite{chong2017programming,beverland2022assessing}. Tasks in simulation and many-body physics include inferring hidden Hamiltonians and open-system dynamics~\cite{wang2017hamiltonian,devega2017dynamics}, while communication and sensing covers, for example, entanglement purification across repeater arrays~\cite{briegel1998repeaters} and GHZ-assisted frequency estimation under non-Markovian noise~\cite{chin2012metrology}. The tasks also cover the full quantum computing stack~\cite{Jones2012layered}, as shown in the  five layers in Fig.~\ref{fig:distribution}b. In our benchmark, the agent needs to reason across different layers.

As shown in Fig.~\ref{fig:distribution}, the tasks run on simulated hardware. Several tasks require no specific
device and are marked device-less. Orthogonally, each task is rated in one
of two modes: i) pass--fail, where a hidden verifier returns a Boolean pass
or fail, or ii) racing, where submissions are
ranked by a Bradley--Terry Elo fit (see Methods). Finally, three deliberately simple test cases
(e.g., measuring the $T_1$ of a transmon qubit) stress-test the benchmark
infrastructure and are excluded from all science scores. We further note that per-task hidden parameters and pass thresholds stay out of all published material, so future systems cannot learn the answers from this paper.

We established three principles during the development of this benchmark. First, every task is designed as an end-to-end laboratory mini-project whose deliverable is a significant final result, such as a physical quantity, a calibrated artifact, or an optimized design. Second, we require that a task's difficulty come from the physics and not from obscurity. Each task has an intended solution path that a competent experimentalist can follow, and a task is hard only when that path turns on a nonobvious inference. Third, to increase the difficulty and mimic the laboratory setting, contributors tend to design traps (or tricks) that can mislead the model. However, we require that every trap be honest, of a kind that arises in realistic experimental environments. For example, stale notebooks are labeled as notebooks and budgets are stated explicitly. The agent is not punished for missing information due to our infrastructure.
\subsection{Benchmark results}\label{sec:benchmark_results}

\begin{figure*}[t]
\centering
\includegraphics[width=\textwidth]{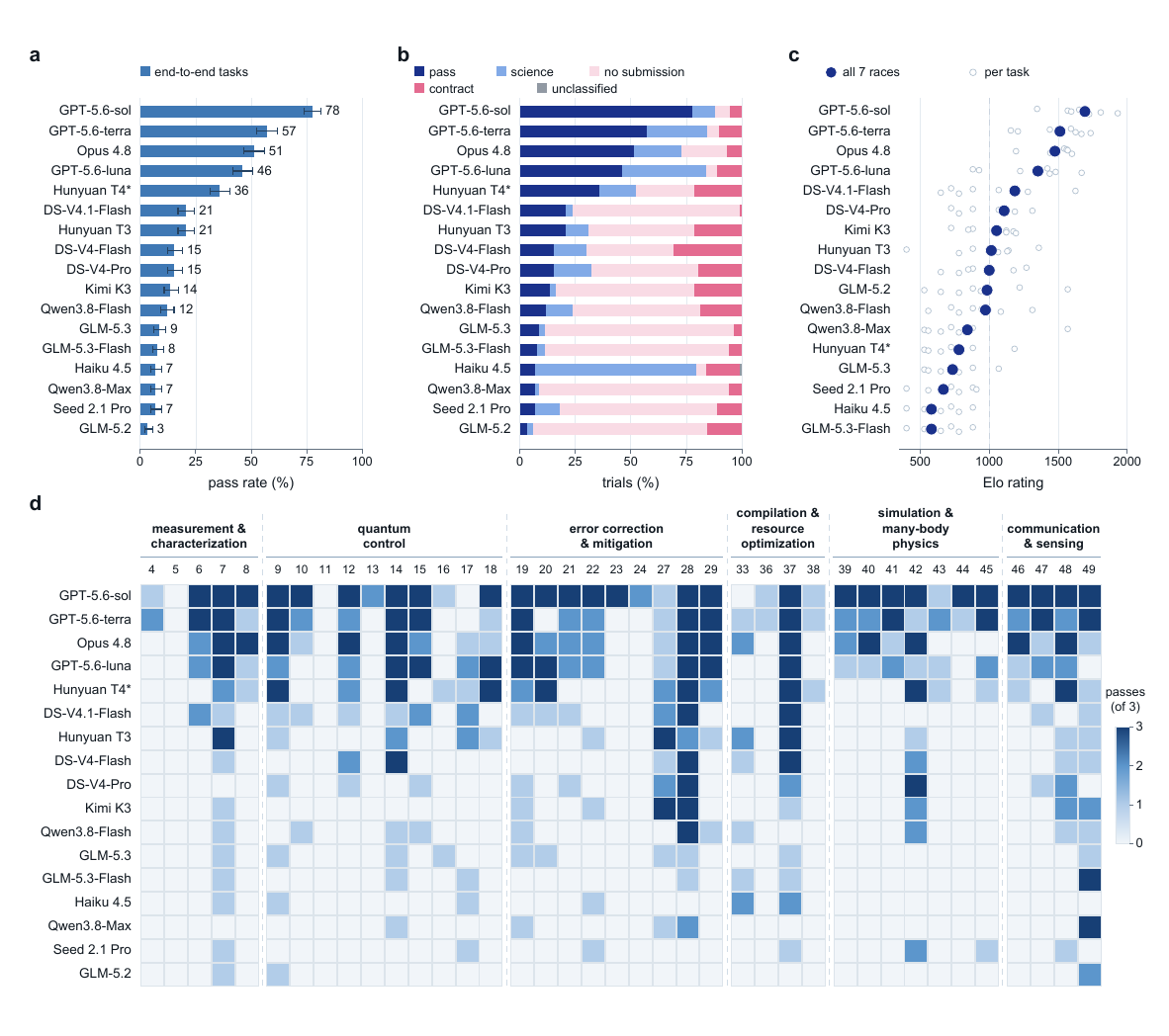}
\caption{Benchmark results and analysis of failure and pass patterns.
\textbf{a.}~Pass rate on the pass--fail tasks. Error bars: Wilson 68\% intervals.
\textbf{b.}~Outcome composition of the trials, including passes, science failures,
contract failures, and no-submission trials.
\textbf{c.}~Bradley--Terry Elo ratings over the 7 racing tasks (filled
circles: games-weighted aggregate; open circles: per-task ratings).
\textbf{d.}~Per-task passes for every system, grouped by the scientific
categories of Fig.~\ref{fig:distribution}, easiest to hardest within each
group. Only pass-fail tasks are shown here. }
\label{fig:results}
\end{figure*}

We evaluated seventeen agentic systems from eight vendors, three from OpenAI
(GPT-5.6-sol, GPT-5.6-terra, and GPT-5.6-luna), three from DeepSeek
(DeepSeek-V4.1-Flash, V4-Pro, and V4-Flash, hereafter
DS-V4.1-Flash and so on), three from Z.ai (GLM-5.3,
5.3-Flash, and 5.2), two from Anthropic (Claude Opus 4.8 and Haiku
4.5), two from Alibaba (Qwen3.8-Max, 3.8-Flash), two from Tencent
(Hunyuan T3 and 4 preview, denoted by Hunyuan T4*), one from Moonshot (Kimi K3), and one
from ByteDance (Seed 2.1 Pro). The two Anthropic systems ran under the Claude
Code agent harness and the three OpenAI systems under the Codex CLI harness.
The remaining twelve systems ran under the OpenCode harness.
Every system ran at its extra high (xhigh) or higher reasoning-effort setting and attempted
each task three times independently. 

Fig.~\ref{fig:results}a ranks the seventeen systems by pass rate on the
pass--fail tasks.
Pass rates run from $78\%$ down to $3\%$ with performance decreasing steadily across the ranked systems and no clear separation into groups.
We note that systems scoring within a narrow band on a frontier agentic benchmark spread across almost the whole range of QIQCBench.
For example, on AutomationBench-AA, the independent run of Zapier's
AutomationBench~\cite{zapier2026automationbench} carried out by Artificial
Analysis~\cite{artificialanalysis2026automationbench}, eight of our seventeen
systems score between $54.0\%$ and $62.2\%$. For GDPval-AA v2~\cite{artificialanalysis2026gdpval}, a widely adopted evaluation, six of them sit within 31 Elo points of one another. We conclude that our benchmark can separate systems that perform alike on other
frontier agentic benchmarks.

We classify failures into three categories (Fig.~\ref{fig:results}b), including no submission when no answer arrives, science failure when the hidden grader rejects the physics, and contract failure when the answer breaks the submission protocol. 
These categories correspond to three requirements of autonomous experimentation: completing the task, reaching a scientifically valid conclusion, and returning that conclusion in a form that can be verified. 
No submission is the largest failure category, accounting for 937 of 1,510 failures. Seven systems exhaust their budget before submitting an answer in more than 60\% of trials. 
These trials are, for the most part, neither crashes nor idle hangs. When timed out, the agents are often still running experiments. Among the remaining failures, $54\%$ are science failures and the rest are contract failures. 
Across the pool, many failures arise from how agents conduct experiments, including relying on stale priors, mismanaging budgets, and discarding in-hand results at submission time.

The racing tasks score performance on a continuous metric, such as required resources or run time (Fig.~\ref{fig:results}c). Every trial races pairwise against the trials of every other system. A trial that clears the task-specific feasibility gate always ranks above one that does not. Among gate-passing trials, the ranking is determined by the recomputed metric (see Methods). 
Of the 357 race trials, 125 cleared their gates. A Bradley--Terry fit spreads the field from 1694 down to 582, with the anchor fixed at 1000. The ordering on the racing tasks corroborates that on the pass--fail tasks, with the same top three systems, but produces wider margins by incorporating design quality alongside reliability.

We examine the pass pattern in Fig.~\ref{fig:results}d. Thirty-seven of the 39 tasks were solved by at least one system. For the two exceptions, a
surrogate modeling on 60 qubits and a net-zero
tunable-coupler CZ gate design, expert reviewers confirmed the intended
solution paths. Six tasks were solved only by the three highest-ranked
systems. Per-task ratings,
gate outcomes, the failure-category definitions, and the audit protocols are
tabulated in the Supplementary Material.

We observe several interesting patterns in the benchmark result. We can observe from Fig.~\ref{fig:results}d that simulation and many-body physics is the hardest category for the pool, with a passing rate of only 15\%.  The top three systems pass 64\% of its trials and the other fourteen pass only 4\%, also the largest such gap of any category. We found no significant correlation between difficulty and stack depth, as the pass rate is uncorrelated with the number of layers a task spans (Spearman $\rho=0.14$, $p=0.40$, see Supplementary Material) and both unsolved tasks sit in the bottom two layers, gates \& calibration and physical device. Success is intermittent, with only 34\% of the 249 system and task pairs that pass at least once passing all three attempts.

\section{Discussion}\label{sec:discussion}
The present benchmark results show that current agents exhibit meaningful but uneven capability on demanding quantum engineering tasks. Large performance gaps between systems and substantial variability across repeated attempts indicate that reliable completion remains a central challenge for autonomous quantum engineering. The failure patterns point to experimental judgment as an important direction for improvement. Choosing informative measurements, revising stale assumptions, and deciding when to stop are recurring demands when agents must reach supported conclusions under finite resources. These failures are not explained by limitations in scientific knowledge or laboratory execution alone, suggesting that improvements to agent harnesses could substantially increase reliability. 

Quantum-Harbor makes the verified autonomy of AI agents measurable by evaluating not only their outputs but also whether those outputs are supported by information acquired during execution. By placing agentic systems in controlled quantum environments, it provides a framework for examining how agents gather evidence, revise hypotheses, allocate experimental resources and arrive at verifiable conclusions. QIQCBench realizes this framework across a broad range of cross-layer quantum engineering tasks, revealing differences between agentic systems that are not apparent from domain-knowledge and coding benchmarks alone. 

Beyond the present benchmark, Quantum-Harbor could support controlled studies of why agents succeed or fail. Task conditions, operating instructions and resource constraints could be varied systematically to disentangle the effects of model capability, harness design and scientific decision-making. Such studies could track progress beyond aggregate pass rates and guide improvements in the reliability and efficiency of autonomous quantum engineering.

We note that the present results are limited to simulated environments and characterize complete agentic systems operating through different harnesses under fixed resource budgets. Future work should determine how these findings transfer across experimental conditions and physical devices, and whether increasingly capable agents can reduce the measurement cost and human effort required for quantum engineering. 
\section{Methods}\label{sec:methods}
\emph{Task construction and review.} Each task follows a common construction and review pipeline.
A domain expert prepares a task-design brief specifying the scientific objective, the intended solution approach, and the task-specific pitfalls. 
A maintainer then builds the task on the Quantum-Harbor runtime, implementing the public materials, the hidden configuration, and the grader as separate artifacts, with the runtime enforcing the public--hidden boundary (Supplemental Material).
Before the task is included in the suite, a second expert reviews the package.
We verify the task end-to-end  on Quantum-Harbor and use coding agents (Claude Code or OpenCode with Fable and Opus 5) to audit selected agents' outcomes as a sanity check.

\emph{Evaluation protocol.} Each trial runs the agent container and the hardware-sidecar container. After the agent's session ends, the sidecar container grades the run. The per-task time budget is 3 hours for all tasks. The agent container has internet access, allowing the system to consult public sources during the run. All tasks ran against the simulation backend. We used coding agents to audit every trial's complete action record. Trials affected by confirmed infrastructure faults were quarantined and rerun afterward.

\emph{Failure categories.} 
We distinguish three failure categories for pass-fail tasks, as stated in the main text. 
The science failure occurs when an admissible submission fails the task's scientific checks.
Depending on the grading contract, the hidden grader checks the submission against hidden ground truth, recomputes reported results from logged evidence, or evaluates the submission in a replay on fresh data. 
The contract failure occurs when a submission is rejected at an admissibility gate before its scientific content is scored, owing to a schema violation, missing mandatory evidence, or delivery outside the published submission channel.
The no-submission trial ends without a final answer being delivered.
Categories were assigned from each failing trial's complete action record, by expert transcript audit or directly from the trial's scorer and termination records, with the source recorded per trial.

\emph{Rating of racing tasks.} Each racing task runs on a fixed instance, so the grader-recomputed race metrics of different systems' submissions are directly comparable.
Every trial is compared pairwise against every trial of every other system on the task.
A trial that passes the task's feasibility gate beats a gate-failing trial. Two gate-passing trials are ordered by the recomputed race metric. Exact ties and pairs of gate failures are excluded from the fit.
Thus, a gate-failing trial contributes a loss against each gate-passing trial of another system, but gate failures are not ranked among themselves.
For the retained comparisons, the Bradley--Terry model assigns each system $i$ a latent strength $\theta_i$ and models the probability that system $i$ beats system $j$ as $\sigma(\theta_i-\theta_j)$, where $\sigma(x)=1/(1+e^{-x})$ is the logistic function.
Let $w_{ij}$ denote the number of retained comparisons won by system $i$
against system $j$. For each task, we fit a Bradley--Terry model by maximizing
$\sum_{i,j} w_{ij}\ln\sigma(\theta_i-\theta_j)
-\tfrac{\lambda}{2}\lVert\theta\rVert^{2}$.
The ridge penalty, with $\lambda=0.5$, keeps the fitted strengths finite
even for systems with only wins or only losses.
We report strengths on the Elo scale,
$\mathrm{Elo}_i = 1000 + (400/\ln 10)(\theta_i-\bar{\theta})$,
where $\bar{\theta}$ is the mean fitted strength across systems rated
on that task. This scaling gives each task a mean rating of 1000.
A 400-point rating advantage corresponds to model-implied win odds
of $10:1$.
A system's aggregate rating is the weighted mean of its per-task ratings, with each task weighted by that system's number of retained comparisons (wins plus losses). Tasks on
which the system has no retained comparisons are excluded from its aggregate.
The rating combines how often a system produces a feasible design with
the quality of its feasible designs.

\emph{Statistics.} A system's pass rate in Fig.~\ref{fig:results}a is $k/n$, the number $k$ of passing trials over its $n=117$ pass--fail trials (39 tasks, three repetitions each). The error bars are Wilson score intervals,
\begin{equation*}
\frac{k+z^{2}/2}{n+z^{2}} \pm \frac{z}{n+z^{2}}
\sqrt{\frac{k(n-k)}{n}+\frac{z^{2}}{4}},
\end{equation*}
with $z=1$, corresponding to a nominal confidence level of approximately $68\%$. These intervals use a pooled binomial approximation and do not quantify uncertainty due to the choice of tasks.
\subsection{Acknowledgement}
\emph{}
The authors thank Feng Pan, Zhan Yu, Yunlong Xiao and Ikko Hamamura for helpful discussions.

%\bibliography{main}

\begin{thebibliography}{66}%
\makeatletter
\providecommand \@ifxundefined [1]{%
 \@ifx{#1\undefined}
}%
\providecommand \@ifnum [1]{%
 \ifnum #1\expandafter \@firstoftwo
 \else \expandafter \@secondoftwo
 \fi
}%
\providecommand \@ifx [1]{%
 \ifx #1\expandafter \@firstoftwo
 \else \expandafter \@secondoftwo
 \fi
}%
\providecommand \natexlab [1]{#1}%
\providecommand \enquote  [1]{``#1''}%
\providecommand \bibnamefont  [1]{#1}%
\providecommand \bibfnamefont [1]{#1}%
\providecommand \citenamefont [1]{#1}%
\providecommand \href@noop [0]{\@secondoftwo}%
\providecommand \href [0]{\begingroup \@sanitize@url \@href}%
\providecommand \@href[1]{\@@startlink{#1}\@@href}%
\providecommand \@@href[1]{\endgroup#1\@@endlink}%
\providecommand \@sanitize@url [0]{\catcode `\\12\catcode `\$12\catcode `\&12\catcode `\#12\catcode `\^12\catcode `\_12\catcode `\%12\relax}%
\providecommand \@@startlink[1]{}%
\providecommand \@@endlink[0]{}%
\providecommand \url  [0]{\begingroup\@sanitize@url \@url }%
\providecommand \@url [1]{\endgroup\@href {#1}{\urlprefix }}%
\providecommand \urlprefix  [0]{URL }%
\providecommand \Eprint [0]{\href }%
\providecommand \doibase [0]{https://doi.org/}%
\providecommand \selectlanguage [0]{\@gobble}%
\providecommand \bibinfo  [0]{\@secondoftwo}%
\providecommand \bibfield  [0]{\@secondoftwo}%
\providecommand \translation [1]{[#1]}%
\providecommand \BibitemOpen [0]{}%
\providecommand \bibitemStop [0]{}%
\providecommand \bibitemNoStop [0]{.\EOS\space}%
\providecommand \EOS [0]{\spacefactor3000\relax}%
\providecommand \BibitemShut  [1]{\csname bibitem#1\endcsname}%
\let\auto@bib@innerbib\@empty
\bibitem [{\citenamefont {Arute}\ \emph {et~al.}(2019)\citenamefont {Arute} \emph {et~al.}}]{arute2019quantum}%
  \BibitemOpen
  \bibfield  {author} {\bibinfo {author} {\bibfnamefont {F.}~\bibnamefont {Arute}} \emph {et~al.},\ }\bibfield  {title} {\bibinfo {title} {Quantum supremacy using a programmable superconducting processor},\ }\href {https://doi.org/10.1038/s41586-019-1666-5} {\bibfield  {journal} {\bibinfo  {journal} {Nature}\ }\textbf {\bibinfo {volume} {574}},\ \bibinfo {pages} {505} (\bibinfo {year} {2019})}\BibitemShut {NoStop}%
\bibitem [{\citenamefont {Daley}\ \emph {et~al.}(2022)\citenamefont {Daley}, \citenamefont {Bloch}, \citenamefont {Kokail}, \citenamefont {Flannigan}, \citenamefont {Pearson}, \citenamefont {Troyer},\ and\ \citenamefont {Zoller}}]{daley2022practical}%
  \BibitemOpen
  \bibfield  {author} {\bibinfo {author} {\bibfnamefont {A.~J.}\ \bibnamefont {Daley}}, \bibinfo {author} {\bibfnamefont {I.}~\bibnamefont {Bloch}}, \bibinfo {author} {\bibfnamefont {C.}~\bibnamefont {Kokail}}, \bibinfo {author} {\bibfnamefont {S.}~\bibnamefont {Flannigan}}, \bibinfo {author} {\bibfnamefont {N.}~\bibnamefont {Pearson}}, \bibinfo {author} {\bibfnamefont {M.}~\bibnamefont {Troyer}},\ and\ \bibinfo {author} {\bibfnamefont {P.}~\bibnamefont {Zoller}},\ }\bibfield  {title} {\bibinfo {title} {Practical quantum advantage in quantum simulation},\ }\href {https://doi.org/10.1038/s41586-022-04940-6} {\bibfield  {journal} {\bibinfo  {journal} {Nature}\ }\textbf {\bibinfo {volume} {607}},\ \bibinfo {pages} {667} (\bibinfo {year} {2022})}\BibitemShut {NoStop}%
\bibitem [{\citenamefont {Azuma}\ \emph {et~al.}(2023)\citenamefont {Azuma}, \citenamefont {Economou}, \citenamefont {Elkouss}, \citenamefont {Hilaire}, \citenamefont {Jiang}, \citenamefont {Lo},\ and\ \citenamefont {Tzitrin}}]{koji2023repeaters}%
  \BibitemOpen
  \bibfield  {author} {\bibinfo {author} {\bibfnamefont {K.}~\bibnamefont {Azuma}}, \bibinfo {author} {\bibfnamefont {S.~E.}\ \bibnamefont {Economou}}, \bibinfo {author} {\bibfnamefont {D.}~\bibnamefont {Elkouss}}, \bibinfo {author} {\bibfnamefont {P.}~\bibnamefont {Hilaire}}, \bibinfo {author} {\bibfnamefont {L.}~\bibnamefont {Jiang}}, \bibinfo {author} {\bibfnamefont {H.-K.}\ \bibnamefont {Lo}},\ and\ \bibinfo {author} {\bibfnamefont {I.}~\bibnamefont {Tzitrin}},\ }\bibfield  {title} {\bibinfo {title} {Quantum repeaters: From quantum networks to the quantum internet},\ }\href {https://doi.org/10.1103/RevModPhys.95.045006} {\bibfield  {journal} {\bibinfo  {journal} {Rev. Mod. Phys.}\ }\textbf {\bibinfo {volume} {95}},\ \bibinfo {pages} {045006} (\bibinfo {year} {2023})}\BibitemShut {NoStop}%
\bibitem [{\citenamefont {Degen}\ \emph {et~al.}(2017)\citenamefont {Degen}, \citenamefont {Reinhard},\ and\ \citenamefont {Cappellaro}}]{degen2017sensing}%
  \BibitemOpen
  \bibfield  {author} {\bibinfo {author} {\bibfnamefont {C.~L.}\ \bibnamefont {Degen}}, \bibinfo {author} {\bibfnamefont {F.}~\bibnamefont {Reinhard}},\ and\ \bibinfo {author} {\bibfnamefont {P.}~\bibnamefont {Cappellaro}},\ }\bibfield  {title} {\bibinfo {title} {Quantum sensing},\ }\href {https://doi.org/10.1103/RevModPhys.89.035002} {\bibfield  {journal} {\bibinfo  {journal} {Rev. Mod. Phys.}\ }\textbf {\bibinfo {volume} {89}},\ \bibinfo {pages} {035002} (\bibinfo {year} {2017})}\BibitemShut {NoStop}%
\bibitem [{\citenamefont {{Google Quantum AI and Collaborators}}(2025{\natexlab{a}})}]{acharya2024threshold}%
  \BibitemOpen
  \bibfield  {author} {\bibinfo {author} {\bibnamefont {{Google Quantum AI and Collaborators}}},\ }\bibfield  {title} {\bibinfo {title} {Quantum error correction below the surface code threshold},\ }\href {https://doi.org/10.1038/s41586-024-08449-y} {\bibfield  {journal} {\bibinfo  {journal} {Nature}\ }\textbf {\bibinfo {volume} {638}},\ \bibinfo {pages} {920} (\bibinfo {year} {2025}{\natexlab{a}})}\BibitemShut {NoStop}%
\bibitem [{\citenamefont {Bluvstein}\ \emph {et~al.}(2024)\citenamefont {Bluvstein} \emph {et~al.}}]{bluvstein2024logical}%
  \BibitemOpen
  \bibfield  {author} {\bibinfo {author} {\bibfnamefont {D.}~\bibnamefont {Bluvstein}} \emph {et~al.},\ }\bibfield  {title} {\bibinfo {title} {Logical quantum processor based on reconfigurable atom arrays},\ }\href {https://doi.org/10.1038/s41586-023-06927-3} {\bibfield  {journal} {\bibinfo  {journal} {Nature}\ }\textbf {\bibinfo {volume} {626}},\ \bibinfo {pages} {58} (\bibinfo {year} {2024})}\BibitemShut {NoStop}%
\bibitem [{\citenamefont {Paetznick}\ \emph {et~al.}(2026)\citenamefont {Paetznick} \emph {et~al.}}]{paetznick2026logical}%
  \BibitemOpen
  \bibfield  {author} {\bibinfo {author} {\bibfnamefont {A.}~\bibnamefont {Paetznick}} \emph {et~al.},\ }\bibfield  {title} {\bibinfo {title} {Improved quantum processor logical error rates via correction and detection},\ }\href {https://doi.org/10.1038/s41586-026-10628-y} {\bibfield  {journal} {\bibinfo  {journal} {Nature}\ }\textbf {\bibinfo {volume} {654}},\ \bibinfo {pages} {349} (\bibinfo {year} {2026})}\BibitemShut {NoStop}%
\bibitem [{\citenamefont {Kim}\ \emph {et~al.}(2023)\citenamefont {Kim} \emph {et~al.}}]{kim2023utility}%
  \BibitemOpen
  \bibfield  {author} {\bibinfo {author} {\bibfnamefont {Y.}~\bibnamefont {Kim}} \emph {et~al.},\ }\bibfield  {title} {\bibinfo {title} {Evidence for the utility of quantum computing before fault tolerance},\ }\href {https://doi.org/10.1038/s41586-023-06096-3} {\bibfield  {journal} {\bibinfo  {journal} {Nature}\ }\textbf {\bibinfo {volume} {618}},\ \bibinfo {pages} {500} (\bibinfo {year} {2023})}\BibitemShut {NoStop}%
\bibitem [{\citenamefont {{Google Quantum AI and Collaborators}}(2025{\natexlab{b}})}]{google2025echoes}%
  \BibitemOpen
  \bibfield  {author} {\bibinfo {author} {\bibnamefont {{Google Quantum AI and Collaborators}}},\ }\bibfield  {title} {\bibinfo {title} {Observation of constructive interference at the edge of quantum ergodicity},\ }\href {https://doi.org/10.1038/s41586-025-09526-6} {\bibfield  {journal} {\bibinfo  {journal} {Nature}\ }\textbf {\bibinfo {volume} {646}},\ \bibinfo {pages} {825} (\bibinfo {year} {2025}{\natexlab{b}})}\BibitemShut {NoStop}%
\bibitem [{\citenamefont {Shaw}\ \emph {et~al.}(2024)\citenamefont {Shaw}, \citenamefont {Chen}, \citenamefont {Choi}, \citenamefont {Mark}, \citenamefont {Scholl}, \citenamefont {Finkelstein}, \citenamefont {Elben}, \citenamefont {Choi},\ and\ \citenamefont {Endres}}]{shaw2024benchmarking}%
  \BibitemOpen
  \bibfield  {author} {\bibinfo {author} {\bibfnamefont {A.~L.}\ \bibnamefont {Shaw}}, \bibinfo {author} {\bibfnamefont {Z.}~\bibnamefont {Chen}}, \bibinfo {author} {\bibfnamefont {J.}~\bibnamefont {Choi}}, \bibinfo {author} {\bibfnamefont {D.~K.}\ \bibnamefont {Mark}}, \bibinfo {author} {\bibfnamefont {P.}~\bibnamefont {Scholl}}, \bibinfo {author} {\bibfnamefont {R.}~\bibnamefont {Finkelstein}}, \bibinfo {author} {\bibfnamefont {A.}~\bibnamefont {Elben}}, \bibinfo {author} {\bibfnamefont {S.}~\bibnamefont {Choi}},\ and\ \bibinfo {author} {\bibfnamefont {M.}~\bibnamefont {Endres}},\ }\bibfield  {title} {\bibinfo {title} {Benchmarking highly entangled states on a 60-atom analogue quantum simulator},\ }\href {https://doi.org/10.1038/s41586-024-07173-x} {\bibfield  {journal} {\bibinfo  {journal} {Nature}\ }\textbf {\bibinfo {volume} {628}},\ \bibinfo {pages} {71} (\bibinfo {year} {2024})}\BibitemShut {NoStop}%
\bibitem [{\citenamefont {Liu}\ \emph {et~al.}(2026)\citenamefont {Liu}, \citenamefont {Zhou}, \citenamefont {Chen}, \citenamefont {Wang}, \citenamefont {Teng}, \citenamefont {Han}, \citenamefont {Liu}, \citenamefont {Zhang}, \citenamefont {Yang}, \citenamefont {Liu}, \citenamefont {Xue}, \citenamefont {Yang}, \citenamefont {Yang}, \citenamefont {Zeng}, \citenamefont {Pan}, \citenamefont {Zheng}, \citenamefont {Zhang}, \citenamefont {Cao}, \citenamefont {Zhen}, \citenamefont {Xiao}, \citenamefont {Li}, \citenamefont {You}, \citenamefont {Ma}, \citenamefont {Zhao}, \citenamefont {Xu}, \citenamefont {Wang}, \citenamefont {Wan}, \citenamefont {Zhang},\ and\ \citenamefont {Pan}}]{liu2026longlived}%
  \BibitemOpen
  \bibfield  {author} {\bibinfo {author} {\bibfnamefont {W.-Z.}\ \bibnamefont {Liu}}, \bibinfo {author} {\bibfnamefont {Y.-B.}\ \bibnamefont {Zhou}}, \bibinfo {author} {\bibfnamefont {J.-P.}\ \bibnamefont {Chen}}, \bibinfo {author} {\bibfnamefont {B.}~\bibnamefont {Wang}}, \bibinfo {author} {\bibfnamefont {A.}~\bibnamefont {Teng}}, \bibinfo {author} {\bibfnamefont {X.-W.}\ \bibnamefont {Han}}, \bibinfo {author} {\bibfnamefont {G.-C.}\ \bibnamefont {Liu}}, \bibinfo {author} {\bibfnamefont {Z.-J.}\ \bibnamefont {Zhang}}, \bibinfo {author} {\bibfnamefont {Y.}~\bibnamefont {Yang}}, \bibinfo {author} {\bibfnamefont {F.-G.}\ \bibnamefont {Liu}}, \bibinfo {author} {\bibfnamefont {C.-H.}\ \bibnamefont {Xue}}, \bibinfo {author} {\bibfnamefont {B.-W.}\ \bibnamefont {Yang}}, \bibinfo {author} {\bibfnamefont {J.}~\bibnamefont {Yang}}, \bibinfo {author} {\bibfnamefont {C.}~\bibnamefont {Zeng}}, \bibinfo {author} {\bibfnamefont {D.-R.}\ \bibnamefont {Pan}}, \bibinfo {author} {\bibfnamefont {M.-Y.}\ \bibnamefont {Zheng}},
  \bibinfo {author} {\bibfnamefont {X.}~\bibnamefont {Zhang}}, \bibinfo {author} {\bibfnamefont {S.}~\bibnamefont {Cao}}, \bibinfo {author} {\bibfnamefont {Y.-Z.}\ \bibnamefont {Zhen}}, \bibinfo {author} {\bibfnamefont {Y.}~\bibnamefont {Xiao}}, \bibinfo {author} {\bibfnamefont {H.}~\bibnamefont {Li}}, \bibinfo {author} {\bibfnamefont {L.}~\bibnamefont {You}}, \bibinfo {author} {\bibfnamefont {X.}~\bibnamefont {Ma}}, \bibinfo {author} {\bibfnamefont {Q.}~\bibnamefont {Zhao}}, \bibinfo {author} {\bibfnamefont {F.}~\bibnamefont {Xu}}, \bibinfo {author} {\bibfnamefont {Y.}~\bibnamefont {Wang}}, \bibinfo {author} {\bibfnamefont {Y.}~\bibnamefont {Wan}}, \bibinfo {author} {\bibfnamefont {Q.}~\bibnamefont {Zhang}},\ and\ \bibinfo {author} {\bibfnamefont {J.-W.}\ \bibnamefont {Pan}},\ }\bibfield  {title} {\bibinfo {title} {Long-lived remote ion–ion entanglement for scalable quantum repeaters},\ }\href {https://doi.org/10.1038/s41586-026-10177-4} {\bibfield  {journal} {\bibinfo  {journal} {Nature}\ }\textbf
  {\bibinfo {volume} {652}},\ \bibinfo {pages} {51} (\bibinfo {year} {2026})}\BibitemShut {NoStop}%
\bibitem [{\citenamefont {Zheng}\ \emph {et~al.}(2026)\citenamefont {Zheng}, \citenamefont {Wang}, \citenamefont {Jia}, \citenamefont {Huang}, \citenamefont {Yuan}, \citenamefont {Zhai}, \citenamefont {Dai}, \citenamefont {Shi}, \citenamefont {Zhang}, \citenamefont {Zhang}, \citenamefont {Zhuang}, \citenamefont {Liu}, \citenamefont {Mao}, \citenamefont {Dai}, \citenamefont {Fu}, \citenamefont {Jiao}, \citenamefont {Shi}, \citenamefont {Dai}, \citenamefont {Wang}, \citenamefont {Li}, \citenamefont {Gong}, \citenamefont {Yuan}, \citenamefont {Chang},\ and\ \citenamefont {Wang}}]{zheng2026largescale}%
  \BibitemOpen
  \bibfield  {author} {\bibinfo {author} {\bibfnamefont {Y.}~\bibnamefont {Zheng}}, \bibinfo {author} {\bibfnamefont {H.}~\bibnamefont {Wang}}, \bibinfo {author} {\bibfnamefont {X.}~\bibnamefont {Jia}}, \bibinfo {author} {\bibfnamefont {J.}~\bibnamefont {Huang}}, \bibinfo {author} {\bibfnamefont {H.}~\bibnamefont {Yuan}}, \bibinfo {author} {\bibfnamefont {C.}~\bibnamefont {Zhai}}, \bibinfo {author} {\bibfnamefont {J.}~\bibnamefont {Dai}}, \bibinfo {author} {\bibfnamefont {J.}~\bibnamefont {Shi}}, \bibinfo {author} {\bibfnamefont {L.}~\bibnamefont {Zhang}}, \bibinfo {author} {\bibfnamefont {X.}~\bibnamefont {Zhang}}, \bibinfo {author} {\bibfnamefont {M.}~\bibnamefont {Zhuang}}, \bibinfo {author} {\bibfnamefont {J.}~\bibnamefont {Liu}}, \bibinfo {author} {\bibfnamefont {J.}~\bibnamefont {Mao}}, \bibinfo {author} {\bibfnamefont {T.}~\bibnamefont {Dai}}, \bibinfo {author} {\bibfnamefont {Z.}~\bibnamefont {Fu}}, \bibinfo {author} {\bibfnamefont {Y.}~\bibnamefont {Jiao}}, \bibinfo {author} {\bibfnamefont
  {Y.}~\bibnamefont {Shi}}, \bibinfo {author} {\bibfnamefont {D.}~\bibnamefont {Dai}}, \bibinfo {author} {\bibfnamefont {X.}~\bibnamefont {Wang}}, \bibinfo {author} {\bibfnamefont {Y.}~\bibnamefont {Li}}, \bibinfo {author} {\bibfnamefont {Q.}~\bibnamefont {Gong}}, \bibinfo {author} {\bibfnamefont {Z.}~\bibnamefont {Yuan}}, \bibinfo {author} {\bibfnamefont {L.}~\bibnamefont {Chang}},\ and\ \bibinfo {author} {\bibfnamefont {J.}~\bibnamefont {Wang}},\ }\bibfield  {title} {\bibinfo {title} {Large-scale quantum communication networks with integrated photonics},\ }\href {https://doi.org/10.1038/s41586-026-10152-z} {\bibfield  {journal} {\bibinfo  {journal} {Nature}\ }\textbf {\bibinfo {volume} {651}},\ \bibinfo {pages} {68} (\bibinfo {year} {2026})}\BibitemShut {NoStop}%
\bibitem [{\citenamefont {Stas}\ \emph {et~al.}(2026)\citenamefont {Stas}, \citenamefont {Wei}, \citenamefont {Sirotin}, \citenamefont {Huan}, \citenamefont {Yazlar}, \citenamefont {Abdo~Arias}, \citenamefont {Knyazev}, \citenamefont {Baranes}, \citenamefont {Machielse}, \citenamefont {Grandi}, \citenamefont {Riedel}, \citenamefont {Borregaard}, \citenamefont {Park}, \citenamefont {Lončar}, \citenamefont {Suleymanzade},\ and\ \citenamefont {Lukin}}]{stas2026entanglement}%
  \BibitemOpen
  \bibfield  {author} {\bibinfo {author} {\bibfnamefont {P.-J.}\ \bibnamefont {Stas}}, \bibinfo {author} {\bibfnamefont {Y.-C.}\ \bibnamefont {Wei}}, \bibinfo {author} {\bibfnamefont {M.}~\bibnamefont {Sirotin}}, \bibinfo {author} {\bibfnamefont {Y.~Q.}\ \bibnamefont {Huan}}, \bibinfo {author} {\bibfnamefont {U.}~\bibnamefont {Yazlar}}, \bibinfo {author} {\bibfnamefont {F.}~\bibnamefont {Abdo~Arias}}, \bibinfo {author} {\bibfnamefont {E.}~\bibnamefont {Knyazev}}, \bibinfo {author} {\bibfnamefont {G.}~\bibnamefont {Baranes}}, \bibinfo {author} {\bibfnamefont {B.}~\bibnamefont {Machielse}}, \bibinfo {author} {\bibfnamefont {S.}~\bibnamefont {Grandi}}, \bibinfo {author} {\bibfnamefont {D.}~\bibnamefont {Riedel}}, \bibinfo {author} {\bibfnamefont {J.}~\bibnamefont {Borregaard}}, \bibinfo {author} {\bibfnamefont {H.}~\bibnamefont {Park}}, \bibinfo {author} {\bibfnamefont {M.}~\bibnamefont {Lončar}}, \bibinfo {author} {\bibfnamefont {A.}~\bibnamefont {Suleymanzade}},\ and\ \bibinfo {author} {\bibfnamefont {M.~D.}\
  \bibnamefont {Lukin}},\ }\bibfield  {title} {\bibinfo {title} {Entanglement-assisted non-local optical interferometry in a quantum network},\ }\href {https://doi.org/10.1038/s41586-026-10171-w} {\bibfield  {journal} {\bibinfo  {journal} {Nature}\ }\textbf {\bibinfo {volume} {651}},\ \bibinfo {pages} {326} (\bibinfo {year} {2026})}\BibitemShut {NoStop}%
\bibitem [{\citenamefont {Zhou}\ \emph {et~al.}(2025)\citenamefont {Zhou}, \citenamefont {Wang}, \citenamefont {Ye}, \citenamefont {Sun}, \citenamefont {Guo}, \citenamefont {Han}, \citenamefont {Chai}, \citenamefont {Ji}, \citenamefont {Xia}, \citenamefont {Shi}, \citenamefont {Wang},\ and\ \citenamefont {Du}}]{zhou2025sensing}%
  \BibitemOpen
  \bibfield  {author} {\bibinfo {author} {\bibfnamefont {X.}~\bibnamefont {Zhou}}, \bibinfo {author} {\bibfnamefont {M.}~\bibnamefont {Wang}}, \bibinfo {author} {\bibfnamefont {X.}~\bibnamefont {Ye}}, \bibinfo {author} {\bibfnamefont {H.}~\bibnamefont {Sun}}, \bibinfo {author} {\bibfnamefont {Y.}~\bibnamefont {Guo}}, \bibinfo {author} {\bibfnamefont {S.}~\bibnamefont {Han}}, \bibinfo {author} {\bibfnamefont {Z.}~\bibnamefont {Chai}}, \bibinfo {author} {\bibfnamefont {W.}~\bibnamefont {Ji}}, \bibinfo {author} {\bibfnamefont {K.}~\bibnamefont {Xia}}, \bibinfo {author} {\bibfnamefont {F.}~\bibnamefont {Shi}}, \bibinfo {author} {\bibfnamefont {Y.}~\bibnamefont {Wang}},\ and\ \bibinfo {author} {\bibfnamefont {J.}~\bibnamefont {Du}},\ }\bibfield  {title} {\bibinfo {title} {Entanglement-enhanced nanoscale single-spin sensing},\ }\href {https://doi.org/10.1038/s41586-025-09790-6} {\bibfield  {journal} {\bibinfo  {journal} {Nature}\ }\textbf {\bibinfo {volume} {647}},\ \bibinfo {pages} {883} (\bibinfo {year}
  {2025})}\BibitemShut {NoStop}%
\bibitem [{\citenamefont {Rovny}\ \emph {et~al.}(2025)\citenamefont {Rovny}, \citenamefont {Kolkowitz},\ and\ \citenamefont {de~Leon}}]{rovny2025multi}%
  \BibitemOpen
  \bibfield  {author} {\bibinfo {author} {\bibfnamefont {J.}~\bibnamefont {Rovny}}, \bibinfo {author} {\bibfnamefont {S.}~\bibnamefont {Kolkowitz}},\ and\ \bibinfo {author} {\bibfnamefont {N.~P.}\ \bibnamefont {de~Leon}},\ }\bibfield  {title} {\bibinfo {title} {Multi-qubit nanoscale sensing with entanglement as a resource},\ }\href {https://doi.org/10.1038/s41586-025-09760-y} {\bibfield  {journal} {\bibinfo  {journal} {Nature}\ }\textbf {\bibinfo {volume} {647}},\ \bibinfo {pages} {876} (\bibinfo {year} {2025})}\BibitemShut {NoStop}%
\bibitem [{\citenamefont {Ji}\ \emph {et~al.}(2024)\citenamefont {Ji}, \citenamefont {Liu}, \citenamefont {Guo}, \citenamefont {Hu}, \citenamefont {Zhou}, \citenamefont {Dai}, \citenamefont {Chen}, \citenamefont {Yu}, \citenamefont {Wang}, \citenamefont {Xia}, \citenamefont {Shi}, \citenamefont {Wang},\ and\ \citenamefont {Du}}]{Ji2024NaturePhotonics}%
  \BibitemOpen
  \bibfield  {author} {\bibinfo {author} {\bibfnamefont {W.}~\bibnamefont {Ji}}, \bibinfo {author} {\bibfnamefont {Z.}~\bibnamefont {Liu}}, \bibinfo {author} {\bibfnamefont {Y.}~\bibnamefont {Guo}}, \bibinfo {author} {\bibfnamefont {Z.}~\bibnamefont {Hu}}, \bibinfo {author} {\bibfnamefont {J.}~\bibnamefont {Zhou}}, \bibinfo {author} {\bibfnamefont {S.}~\bibnamefont {Dai}}, \bibinfo {author} {\bibfnamefont {Y.}~\bibnamefont {Chen}}, \bibinfo {author} {\bibfnamefont {P.}~\bibnamefont {Yu}}, \bibinfo {author} {\bibfnamefont {M.}~\bibnamefont {Wang}}, \bibinfo {author} {\bibfnamefont {K.}~\bibnamefont {Xia}}, \bibinfo {author} {\bibfnamefont {F.}~\bibnamefont {Shi}}, \bibinfo {author} {\bibfnamefont {Y.}~\bibnamefont {Wang}},\ and\ \bibinfo {author} {\bibfnamefont {J.}~\bibnamefont {Du}},\ }\bibfield  {title} {\bibinfo {title} {Correlated sensing with a solid-state quantum multisensor system for atomic-scale structural analysis},\ }\href {https://doi.org/10.1038/s41566-023-01352-4} {\bibfield  {journal} {\bibinfo
   {journal} {Nat. Photonics}\ }\textbf {\bibinfo {volume} {18}},\ \bibinfo {pages} {230} (\bibinfo {year} {2024})}\BibitemShut {NoStop}%
\bibitem [{\citenamefont {Alexeev}\ \emph {et~al.}(2025)\citenamefont {Alexeev}, \citenamefont {Farag}, \citenamefont {Patti}, \citenamefont {Wolf}, \citenamefont {Ares}, \citenamefont {Aspuru-Guzik}, \citenamefont {Benjamin}, \citenamefont {Cai}, \citenamefont {Cao}, \citenamefont {Chamberland} \emph {et~al.}}]{alexeev2025artificial}%
  \BibitemOpen
  \bibfield  {author} {\bibinfo {author} {\bibfnamefont {Y.}~\bibnamefont {Alexeev}}, \bibinfo {author} {\bibfnamefont {M.~H.}\ \bibnamefont {Farag}}, \bibinfo {author} {\bibfnamefont {T.~L.}\ \bibnamefont {Patti}}, \bibinfo {author} {\bibfnamefont {M.~E.}\ \bibnamefont {Wolf}}, \bibinfo {author} {\bibfnamefont {N.}~\bibnamefont {Ares}}, \bibinfo {author} {\bibfnamefont {A.}~\bibnamefont {Aspuru-Guzik}}, \bibinfo {author} {\bibfnamefont {S.~C.}\ \bibnamefont {Benjamin}}, \bibinfo {author} {\bibfnamefont {Z.}~\bibnamefont {Cai}}, \bibinfo {author} {\bibfnamefont {S.}~\bibnamefont {Cao}}, \bibinfo {author} {\bibfnamefont {C.}~\bibnamefont {Chamberland}}, \emph {et~al.},\ }\bibfield  {title} {\bibinfo {title} {Artificial intelligence for quantum computing},\ }\href {https://doi.org/10.1038/s41467-025-65836-3} {\bibfield  {journal} {\bibinfo  {journal} {Nat. Commun.}\ }\textbf {\bibinfo {volume} {16}},\ \bibinfo {pages} {10829} (\bibinfo {year} {2025})}\BibitemShut {NoStop}%
\bibitem [{\citenamefont {Beverland}\ \emph {et~al.}(2022)\citenamefont {Beverland} \emph {et~al.}}]{beverland2022assessing}%
  \BibitemOpen
  \bibfield  {author} {\bibinfo {author} {\bibfnamefont {M.~E.}\ \bibnamefont {Beverland}} \emph {et~al.},\ }\href {https://doi.org/10.48550/arXiv.2211.07629} {\bibinfo {title} {Assessing requirements to scale to practical quantum advantage}} (\bibinfo {year} {2022}),\ \Eprint {https://arxiv.org/abs/2211.07629} {arXiv:2211.07629} \BibitemShut {NoStop}%
\bibitem [{\citenamefont {Mohseni}\ \emph {et~al.}(2024)\citenamefont {Mohseni} \emph {et~al.}}]{mohseni2024supercomputer}%
  \BibitemOpen
  \bibfield  {author} {\bibinfo {author} {\bibfnamefont {M.}~\bibnamefont {Mohseni}} \emph {et~al.},\ }\href {https://doi.org/10.48550/arXiv.2411.10406} {\bibinfo {title} {How to build a quantum supercomputer: Scaling from hundreds to millions of qubits}} (\bibinfo {year} {2024}),\ \bibinfo {note} {version 3, revised March 2026},\ \Eprint {https://arxiv.org/abs/2411.10406} {arXiv:2411.10406} \BibitemShut {NoStop}%
\bibitem [{\citenamefont {Moon}\ \emph {et~al.}(2020)\citenamefont {Moon} \emph {et~al.}}]{moon2020machine}%
  \BibitemOpen
  \bibfield  {author} {\bibinfo {author} {\bibfnamefont {H.}~\bibnamefont {Moon}} \emph {et~al.},\ }\bibfield  {title} {\bibinfo {title} {Machine learning enables completely automatic tuning of a quantum device faster than human experts},\ }\href {https://doi.org/10.1038/s41467-020-17835-9} {\bibfield  {journal} {\bibinfo  {journal} {Nat. Commun.}\ }\textbf {\bibinfo {volume} {11}},\ \bibinfo {pages} {4161} (\bibinfo {year} {2020})}\BibitemShut {NoStop}%
\bibitem [{\citenamefont {Chong}\ \emph {et~al.}(2017)\citenamefont {Chong}, \citenamefont {Franklin},\ and\ \citenamefont {Martonosi}}]{chong2017programming}%
  \BibitemOpen
  \bibfield  {author} {\bibinfo {author} {\bibfnamefont {F.~T.}\ \bibnamefont {Chong}}, \bibinfo {author} {\bibfnamefont {D.}~\bibnamefont {Franklin}},\ and\ \bibinfo {author} {\bibfnamefont {M.}~\bibnamefont {Martonosi}},\ }\bibfield  {title} {\bibinfo {title} {Programming languages and compiler design for realistic quantum hardware},\ }\href {https://doi.org/10.1038/nature23459} {\bibfield  {journal} {\bibinfo  {journal} {Nature}\ }\textbf {\bibinfo {volume} {549}},\ \bibinfo {pages} {180} (\bibinfo {year} {2017})}\BibitemShut {NoStop}%
\bibitem [{\citenamefont {Wang}\ \emph {et~al.}(2017)\citenamefont {Wang}, \citenamefont {Paesani}, \citenamefont {Santagati}, \citenamefont {Knauer}, \citenamefont {Gentile}, \citenamefont {Wiebe}, \citenamefont {Petruzzella}, \citenamefont {O'Brien}, \citenamefont {Rarity}, \citenamefont {Laing},\ and\ \citenamefont {Thompson}}]{wang2017hamiltonian}%
  \BibitemOpen
  \bibfield  {author} {\bibinfo {author} {\bibfnamefont {J.}~\bibnamefont {Wang}}, \bibinfo {author} {\bibfnamefont {S.}~\bibnamefont {Paesani}}, \bibinfo {author} {\bibfnamefont {R.}~\bibnamefont {Santagati}}, \bibinfo {author} {\bibfnamefont {S.}~\bibnamefont {Knauer}}, \bibinfo {author} {\bibfnamefont {A.~A.}\ \bibnamefont {Gentile}}, \bibinfo {author} {\bibfnamefont {N.}~\bibnamefont {Wiebe}}, \bibinfo {author} {\bibfnamefont {M.}~\bibnamefont {Petruzzella}}, \bibinfo {author} {\bibfnamefont {J.~L.}\ \bibnamefont {O'Brien}}, \bibinfo {author} {\bibfnamefont {J.~G.}\ \bibnamefont {Rarity}}, \bibinfo {author} {\bibfnamefont {A.}~\bibnamefont {Laing}},\ and\ \bibinfo {author} {\bibfnamefont {M.~G.}\ \bibnamefont {Thompson}},\ }\bibfield  {title} {\bibinfo {title} {Experimental quantum {Hamiltonian} learning},\ }\href {https://doi.org/10.1038/nphys4074} {\bibfield  {journal} {\bibinfo  {journal} {Nat. Phys.}\ }\textbf {\bibinfo {volume} {13}},\ \bibinfo {pages} {551} (\bibinfo {year} {2017})}\BibitemShut
  {NoStop}%
\bibitem [{\citenamefont {Du}\ \emph {et~al.}(2026)\citenamefont {Du}, \citenamefont {Zhu}, \citenamefont {Zhang}, \citenamefont {Hsieh}, \citenamefont {Rebentrost}, \citenamefont {Gao}, \citenamefont {You}, \citenamefont {Eisert}, \citenamefont {Chiribella}, \citenamefont {Tao}, \citenamefont {Sanders},\ and\ \citenamefont {Wu}}]{du2026artificial}%
  \BibitemOpen
  \bibfield  {author} {\bibinfo {author} {\bibfnamefont {Y.}~\bibnamefont {Du}}, \bibinfo {author} {\bibfnamefont {Y.}~\bibnamefont {Zhu}}, \bibinfo {author} {\bibfnamefont {Y.-H.}\ \bibnamefont {Zhang}}, \bibinfo {author} {\bibfnamefont {M.-H.}\ \bibnamefont {Hsieh}}, \bibinfo {author} {\bibfnamefont {P.}~\bibnamefont {Rebentrost}}, \bibinfo {author} {\bibfnamefont {W.}~\bibnamefont {Gao}}, \bibinfo {author} {\bibfnamefont {Y.-Z.}\ \bibnamefont {You}}, \bibinfo {author} {\bibfnamefont {J.}~\bibnamefont {Eisert}}, \bibinfo {author} {\bibfnamefont {G.}~\bibnamefont {Chiribella}}, \bibinfo {author} {\bibfnamefont {D.}~\bibnamefont {Tao}}, \bibinfo {author} {\bibfnamefont {B.~C.}\ \bibnamefont {Sanders}},\ and\ \bibinfo {author} {\bibfnamefont {Y.-D.}\ \bibnamefont {Wu}},\ }\bibfield  {title} {\bibinfo {title} {Artificial intelligence for representing and characterizing quantum systems},\ }\href {https://doi.org/10.1038/s42254-026-00962-5} {\bibfield  {journal} {\bibinfo  {journal} {Nat. Rev. Phys.}\ }\textbf
  {\bibinfo {volume} {8}},\ \bibinfo {pages} {579} (\bibinfo {year} {2026})}\BibitemShut {NoStop}%
\bibitem [{\citenamefont {Yao}\ \emph {et~al.}(2023)\citenamefont {Yao}, \citenamefont {Zhao}, \citenamefont {Yu}, \citenamefont {Du}, \citenamefont {Shafran}, \citenamefont {Narasimhan},\ and\ \citenamefont {Cao}}]{yao2023react}%
  \BibitemOpen
  \bibfield  {author} {\bibinfo {author} {\bibfnamefont {S.}~\bibnamefont {Yao}}, \bibinfo {author} {\bibfnamefont {J.}~\bibnamefont {Zhao}}, \bibinfo {author} {\bibfnamefont {D.}~\bibnamefont {Yu}}, \bibinfo {author} {\bibfnamefont {N.}~\bibnamefont {Du}}, \bibinfo {author} {\bibfnamefont {I.}~\bibnamefont {Shafran}}, \bibinfo {author} {\bibfnamefont {K.~R.}\ \bibnamefont {Narasimhan}},\ and\ \bibinfo {author} {\bibfnamefont {Y.}~\bibnamefont {Cao}},\ }\bibfield  {title} {\bibinfo {title} {{ReAct}: Synergizing reasoning and acting in language models},\ }in\ \href {https://openreview.net/forum?id=WE_vluYUL-X} {\emph {\bibinfo {booktitle} {The Eleventh International Conference on Learning Representations}}}\ (\bibinfo {year} {2023})\BibitemShut {NoStop}%
\bibitem [{\citenamefont {Cao}\ \emph {et~al.}(2025)\citenamefont {Cao}, \citenamefont {Zhang}, \citenamefont {Alghadeer}, \citenamefont {Fasciati}, \citenamefont {Piscitelli}, \citenamefont {Bakr}, \citenamefont {Leek},\ and\ \citenamefont {Aspuru-Guzik}}]{cao2025kagents}%
  \BibitemOpen
  \bibfield  {author} {\bibinfo {author} {\bibfnamefont {S.}~\bibnamefont {Cao}}, \bibinfo {author} {\bibfnamefont {Z.}~\bibnamefont {Zhang}}, \bibinfo {author} {\bibfnamefont {M.}~\bibnamefont {Alghadeer}}, \bibinfo {author} {\bibfnamefont {S.~D.}\ \bibnamefont {Fasciati}}, \bibinfo {author} {\bibfnamefont {M.}~\bibnamefont {Piscitelli}}, \bibinfo {author} {\bibfnamefont {M.}~\bibnamefont {Bakr}}, \bibinfo {author} {\bibfnamefont {P.}~\bibnamefont {Leek}},\ and\ \bibinfo {author} {\bibfnamefont {A.}~\bibnamefont {Aspuru-Guzik}},\ }\bibfield  {title} {\bibinfo {title} {Automating quantum computing laboratory experiments with an agent-based {AI} framework},\ }\href {https://doi.org/10.1016/j.patter.2025.101372} {\bibfield  {journal} {\bibinfo  {journal} {Patterns}\ }\textbf {\bibinfo {volume} {6}},\ \bibinfo {pages} {101372} (\bibinfo {year} {2025})}\BibitemShut {NoStop}%
\bibitem [{\citenamefont {Xu}\ \emph {et~al.}(2026)\citenamefont {Xu}, \citenamefont {Han}, \citenamefont {Ou} \emph {et~al.}}]{xu2026vibecal}%
  \BibitemOpen
  \bibfield  {author} {\bibinfo {author} {\bibfnamefont {H.}~\bibnamefont {Xu}}, \bibinfo {author} {\bibfnamefont {J.}~\bibnamefont {Han}}, \bibinfo {author} {\bibfnamefont {S.}~\bibnamefont {Ou}}, \emph {et~al.},\ }\href {https://doi.org/10.48550/arXiv.2606.22376} {\bibinfo {title} {Vibe calibration: Autonomous bring-up of a 112-qubit superconducting quantum processor by a skill-orchestrating language agent}} (\bibinfo {year} {2026}),\ \Eprint {https://arxiv.org/abs/2606.22376} {arXiv:2606.22376} \BibitemShut {NoStop}%
\bibitem [{\citenamefont {Arlt}\ \emph {et~al.}(2025)\citenamefont {Arlt}, \citenamefont {Gu},\ and\ \citenamefont {Krenn}}]{arlt2025agents}%
  \BibitemOpen
  \bibfield  {author} {\bibinfo {author} {\bibfnamefont {S.}~\bibnamefont {Arlt}}, \bibinfo {author} {\bibfnamefont {X.}~\bibnamefont {Gu}},\ and\ \bibinfo {author} {\bibfnamefont {M.}~\bibnamefont {Krenn}},\ }\href {https://doi.org/10.48550/arXiv.2511.11752} {\bibinfo {title} {Towards autonomous quantum physics research using {LLM} agents with access to intelligent tools}} (\bibinfo {year} {2025}),\ \Eprint {https://arxiv.org/abs/2511.11752} {arXiv:2511.11752} \BibitemShut {NoStop}%
\bibitem [{\citenamefont {Isogawa}\ \emph {et~al.}(2026)\citenamefont {Isogawa}, \citenamefont {Okabe}, \citenamefont {Phadetsuwannukun}, \citenamefont {Li},\ and\ \citenamefont {Cappellaro}}]{isogawa2026sensing}%
  \BibitemOpen
  \bibfield  {author} {\bibinfo {author} {\bibfnamefont {T.}~\bibnamefont {Isogawa}}, \bibinfo {author} {\bibfnamefont {R.}~\bibnamefont {Okabe}}, \bibinfo {author} {\bibfnamefont {N.}~\bibnamefont {Phadetsuwannukun}}, \bibinfo {author} {\bibfnamefont {M.}~\bibnamefont {Li}},\ and\ \bibinfo {author} {\bibfnamefont {P.}~\bibnamefont {Cappellaro}},\ }\href {https://arxiv.org/abs/2607.25145} {\bibinfo {title} {Agentic {AI} for scientific reasoning in autonomous quantum sensing experiments}} (\bibinfo {year} {2026}),\ \Eprint {https://arxiv.org/abs/2607.25145} {arXiv:2607.25145 [quant-ph]} \BibitemShut {NoStop}%
\bibitem [{\citenamefont {Dalyac}\ \emph {et~al.}(2026)\citenamefont {Dalyac}, \citenamefont {Dauphin}, \citenamefont {Henriet},\ and\ \citenamefont {Jurczak}}]{dalyac2026lowering}%
  \BibitemOpen
  \bibfield  {author} {\bibinfo {author} {\bibfnamefont {C.}~\bibnamefont {Dalyac}}, \bibinfo {author} {\bibfnamefont {A.}~\bibnamefont {Dauphin}}, \bibinfo {author} {\bibfnamefont {L.}~\bibnamefont {Henriet}},\ and\ \bibinfo {author} {\bibfnamefont {C.}~\bibnamefont {Jurczak}},\ }\href@noop {} {\bibinfo {title} {Lowering the implementation barrier of neutral-atom quantum computing with agentic workflows}} (\bibinfo {year} {2026}),\ \Eprint {https://arxiv.org/abs/2607.25834} {arXiv:2607.25834} \BibitemShut {NoStop}%
\bibitem [{\citenamefont {Fu}\ \emph {et~al.}(2025)\citenamefont {Fu}, \citenamefont {Jiang}, \citenamefont {Xu}, \citenamefont {Huang},\ and\ \citenamefont {Chen}}]{fu2025qagent}%
  \BibitemOpen
  \bibfield  {author} {\bibinfo {author} {\bibfnamefont {Z.}~\bibnamefont {Fu}}, \bibinfo {author} {\bibfnamefont {L.}~\bibnamefont {Jiang}}, \bibinfo {author} {\bibfnamefont {Y.}~\bibnamefont {Xu}}, \bibinfo {author} {\bibfnamefont {G.}~\bibnamefont {Huang}},\ and\ \bibinfo {author} {\bibfnamefont {F.}~\bibnamefont {Chen}},\ }\href {https://arxiv.org/abs/2508.20134} {\bibinfo {title} {{QAgent}: An {LLM}-based multi-agent system for autonomous {OpenQASM} programming}} (\bibinfo {year} {2025}),\ \Eprint {https://arxiv.org/abs/2508.20134} {arXiv:2508.20134 [cs.AI]} \BibitemShut {NoStop}%
\bibitem [{\citenamefont {Li}\ \emph {et~al.}(2026)\citenamefont {Li}, \citenamefont {Ren}, \citenamefont {Cheng},\ and\ \citenamefont {Gong}}]{li2026autosim}%
  \BibitemOpen
  \bibfield  {author} {\bibinfo {author} {\bibfnamefont {W.}~\bibnamefont {Li}}, \bibinfo {author} {\bibfnamefont {J.}~\bibnamefont {Ren}}, \bibinfo {author} {\bibfnamefont {L.}~\bibnamefont {Cheng}},\ and\ \bibinfo {author} {\bibfnamefont {C.}~\bibnamefont {Gong}},\ }\href {https://arxiv.org/abs/2601.10194} {\bibinfo {title} {Autonomous quantum simulation through large language model agents}} (\bibinfo {year} {2026}),\ \Eprint {https://arxiv.org/abs/2601.10194} {arXiv:2601.10194 [quant-ph]} \BibitemShut {NoStop}%
\bibitem [{\citenamefont {Gustin}\ \emph {et~al.}(2026)\citenamefont {Gustin}, \citenamefont {Mantilla~Calder{\'o}n}, \citenamefont {P{\'e}rez-S{\'a}nchez}, \citenamefont {Crebolder}, \citenamefont {Gonthier}, \citenamefont {Ghazi~Vakili}, \citenamefont {Nakamura}, \citenamefont {Panicker}, \citenamefont {Ramprasad}, \citenamefont {Yin} \emph {et~al.}}]{gustin2026agente}%
  \BibitemOpen
  \bibfield  {author} {\bibinfo {author} {\bibfnamefont {I.}~\bibnamefont {Gustin}}, \bibinfo {author} {\bibfnamefont {L.}~\bibnamefont {Mantilla~Calder{\'o}n}}, \bibinfo {author} {\bibfnamefont {J.~B.}\ \bibnamefont {P{\'e}rez-S{\'a}nchez}}, \bibinfo {author} {\bibfnamefont {C.}~\bibnamefont {Crebolder}}, \bibinfo {author} {\bibfnamefont {J.~F.}\ \bibnamefont {Gonthier}}, \bibinfo {author} {\bibfnamefont {M.}~\bibnamefont {Ghazi~Vakili}}, \bibinfo {author} {\bibfnamefont {Y.}~\bibnamefont {Nakamura}}, \bibinfo {author} {\bibfnamefont {K.}~\bibnamefont {Panicker}}, \bibinfo {author} {\bibfnamefont {M.}~\bibnamefont {Ramprasad}}, \bibinfo {author} {\bibfnamefont {A.}~\bibnamefont {Yin}}, \emph {et~al.},\ }\bibfield  {title} {\bibinfo {title} {El agente cu{\'a}ntico: automating quantum simulations},\ }\href {https://doi.org/10.1088/1361-6633/ae8933} {\bibfield  {journal} {\bibinfo  {journal} {Rep. Prog. Phys.}\ }\textbf {\bibinfo {volume} {89}},\ \bibinfo {pages} {077602} (\bibinfo {year} {2026})}\BibitemShut
  {NoStop}%
\bibitem [{\citenamefont {Shiraishi}\ \emph {et~al.}(2026)\citenamefont {Shiraishi}, \citenamefont {Hamamura}, \citenamefont {Ishigaki},\ and\ \citenamefont {Kadowaki}}]{shiraishi2026model}%
  \BibitemOpen
  \bibfield  {author} {\bibinfo {author} {\bibfnamefont {M.}~\bibnamefont {Shiraishi}}, \bibinfo {author} {\bibfnamefont {I.}~\bibnamefont {Hamamura}}, \bibinfo {author} {\bibfnamefont {T.}~\bibnamefont {Ishigaki}},\ and\ \bibinfo {author} {\bibfnamefont {T.}~\bibnamefont {Kadowaki}},\ }\bibfield  {title} {\bibinfo {title} {A {Model Context Protocol} server for quantum execution in hybrid quantum-{HPC} environments},\ }in\ \href {https://doi.org/10.1109/QCNC69040.2026.00145} {\emph {\bibinfo {booktitle} {2026 International Conference on Quantum Communications, Networking, and Computing (QCNC)}}}\ (\bibinfo {organization} {IEEE},\ \bibinfo {year} {2026})\ pp.\ \bibinfo {pages} {1--6}\BibitemShut {NoStop}%
\bibitem [{\citenamefont {Proctor}\ \emph {et~al.}(2020)\citenamefont {Proctor}, \citenamefont {Revelle}, \citenamefont {Nielsen}, \citenamefont {Rudinger}, \citenamefont {Lobser}, \citenamefont {Maunz}, \citenamefont {Blume-Kohout},\ and\ \citenamefont {Young}}]{proctor2020drift}%
  \BibitemOpen
  \bibfield  {author} {\bibinfo {author} {\bibfnamefont {T.}~\bibnamefont {Proctor}}, \bibinfo {author} {\bibfnamefont {M.}~\bibnamefont {Revelle}}, \bibinfo {author} {\bibfnamefont {E.}~\bibnamefont {Nielsen}}, \bibinfo {author} {\bibfnamefont {K.}~\bibnamefont {Rudinger}}, \bibinfo {author} {\bibfnamefont {D.}~\bibnamefont {Lobser}}, \bibinfo {author} {\bibfnamefont {P.}~\bibnamefont {Maunz}}, \bibinfo {author} {\bibfnamefont {R.}~\bibnamefont {Blume-Kohout}},\ and\ \bibinfo {author} {\bibfnamefont {K.}~\bibnamefont {Young}},\ }\bibfield  {title} {\bibinfo {title} {Detecting and tracking drift in quantum information processors},\ }\href {https://doi.org/10.1038/s41467-020-19074-4} {\bibfield  {journal} {\bibinfo  {journal} {Nat. Commun.}\ }\textbf {\bibinfo {volume} {11}},\ \bibinfo {pages} {5396} (\bibinfo {year} {2020})}\BibitemShut {NoStop}%
\bibitem [{\citenamefont {Sivak}\ \emph {et~al.}(2026)\citenamefont {Sivak}, \citenamefont {Morvan}, \citenamefont {Broughton}, \citenamefont {Corti{\~n}as}, \citenamefont {Bausch}, \citenamefont {Senior}, \citenamefont {Neeley}, \citenamefont {Eickbusch}, \citenamefont {Shutty}, \citenamefont {Beni} \emph {et~al.}}]{sivak2026reinforcement}%
  \BibitemOpen
  \bibfield  {author} {\bibinfo {author} {\bibfnamefont {V.}~\bibnamefont {Sivak}}, \bibinfo {author} {\bibfnamefont {A.}~\bibnamefont {Morvan}}, \bibinfo {author} {\bibfnamefont {M.}~\bibnamefont {Broughton}}, \bibinfo {author} {\bibfnamefont {R.~G.}\ \bibnamefont {Corti{\~n}as}}, \bibinfo {author} {\bibfnamefont {J.}~\bibnamefont {Bausch}}, \bibinfo {author} {\bibfnamefont {A.~W.}\ \bibnamefont {Senior}}, \bibinfo {author} {\bibfnamefont {M.}~\bibnamefont {Neeley}}, \bibinfo {author} {\bibfnamefont {A.}~\bibnamefont {Eickbusch}}, \bibinfo {author} {\bibfnamefont {N.}~\bibnamefont {Shutty}}, \bibinfo {author} {\bibfnamefont {L.~A.}\ \bibnamefont {Beni}}, \emph {et~al.},\ }\bibfield  {title} {\bibinfo {title} {Reinforcement learning control of quantum error correction},\ }\href@noop {} {\bibfield  {journal} {\bibinfo  {journal} {Nature}\ }\textbf {\bibinfo {volume} {655}},\ \bibinfo {pages} {879–884} (\bibinfo {year} {2026})}\BibitemShut {NoStop}%
\bibitem [{\citenamefont {Lennon}\ \emph {et~al.}(2019)\citenamefont {Lennon}, \citenamefont {Moon}, \citenamefont {Camenzind}, \citenamefont {Yu}, \citenamefont {Zumb{\"u}hl}, \citenamefont {Briggs}, \citenamefont {Osborne}, \citenamefont {Laird},\ and\ \citenamefont {Ares}}]{lennon2019efficiently}%
  \BibitemOpen
  \bibfield  {author} {\bibinfo {author} {\bibfnamefont {D.~T.}\ \bibnamefont {Lennon}}, \bibinfo {author} {\bibfnamefont {H.}~\bibnamefont {Moon}}, \bibinfo {author} {\bibfnamefont {L.~C.}\ \bibnamefont {Camenzind}}, \bibinfo {author} {\bibfnamefont {L.}~\bibnamefont {Yu}}, \bibinfo {author} {\bibfnamefont {D.~M.}\ \bibnamefont {Zumb{\"u}hl}}, \bibinfo {author} {\bibfnamefont {G.~A. .~D.}\ \bibnamefont {Briggs}}, \bibinfo {author} {\bibfnamefont {M.~A.}\ \bibnamefont {Osborne}}, \bibinfo {author} {\bibfnamefont {E.~A.}\ \bibnamefont {Laird}},\ and\ \bibinfo {author} {\bibfnamefont {N.}~\bibnamefont {Ares}},\ }\bibfield  {title} {\bibinfo {title} {Efficiently measuring a quantum device using machine learning},\ }\href {https://doi.org/10.1038/s41534-019-0193-4} {\bibfield  {journal} {\bibinfo  {journal} {npj Quantum Information}\ }\textbf {\bibinfo {volume} {5}},\ \bibinfo {pages} {79} (\bibinfo {year} {2019})}\BibitemShut {NoStop}%
\bibitem [{\citenamefont {Baum}\ \emph {et~al.}(2021)\citenamefont {Baum} \emph {et~al.}}]{baum2021experimental}%
  \BibitemOpen
  \bibfield  {author} {\bibinfo {author} {\bibfnamefont {Y.}~\bibnamefont {Baum}} \emph {et~al.},\ }\bibfield  {title} {\bibinfo {title} {Experimental deep reinforcement learning for error-robust gate-set design on a superconducting quantum computer},\ }\href {https://doi.org/10.1103/PRXQuantum.2.040324} {\bibfield  {journal} {\bibinfo  {journal} {PRX Quantum}\ }\textbf {\bibinfo {volume} {2}},\ \bibinfo {pages} {040324} (\bibinfo {year} {2021})}\BibitemShut {NoStop}%
\bibitem [{\citenamefont {Cao}\ \emph {et~al.}(2026)\citenamefont {Cao} \emph {et~al.}}]{cao2026qcaleval}%
  \BibitemOpen
  \bibfield  {author} {\bibinfo {author} {\bibfnamefont {S.}~\bibnamefont {Cao}} \emph {et~al.},\ }\href@noop {} {\bibinfo {title} {{QCalEval}: Benchmarking vision-language models for quantum calibration plot understanding}} (\bibinfo {year} {2026}),\ \Eprint {https://arxiv.org/abs/2604.25884} {arXiv:2604.25884} \BibitemShut {NoStop}%
\bibitem [{\citenamefont {Minami}\ \emph {et~al.}(2025)\citenamefont {Minami}, \citenamefont {Ishigaki}, \citenamefont {Hamamura} \emph {et~al.}}]{minami2025quantumbench}%
  \BibitemOpen
  \bibfield  {author} {\bibinfo {author} {\bibfnamefont {S.}~\bibnamefont {Minami}}, \bibinfo {author} {\bibfnamefont {T.}~\bibnamefont {Ishigaki}}, \bibinfo {author} {\bibfnamefont {I.}~\bibnamefont {Hamamura}}, \emph {et~al.},\ }\href {https://doi.org/10.48550/arXiv.2511.00092} {\bibinfo {title} {{QuantumBench}: A benchmark for quantum problem solving}} (\bibinfo {year} {2025}),\ \Eprint {https://arxiv.org/abs/2511.00092} {arXiv:2511.00092} \BibitemShut {NoStop}%
\bibitem [{\citenamefont {Yang}\ \emph {et~al.}(2025)\citenamefont {Yang}, \citenamefont {Wang}, \citenamefont {Gu}, \citenamefont {Liang},\ and\ \citenamefont {Li}}]{yang2024qcircuitbench}%
  \BibitemOpen
  \bibfield  {author} {\bibinfo {author} {\bibfnamefont {R.}~\bibnamefont {Yang}}, \bibinfo {author} {\bibfnamefont {Z.}~\bibnamefont {Wang}}, \bibinfo {author} {\bibfnamefont {Y.}~\bibnamefont {Gu}}, \bibinfo {author} {\bibfnamefont {Y.}~\bibnamefont {Liang}},\ and\ \bibinfo {author} {\bibfnamefont {T.}~\bibnamefont {Li}},\ }\bibfield  {title} {\bibinfo {title} {{QCircuitBench}: A large-scale dataset for benchmarking quantum algorithm design},\ }in\ \href {https://doi.org/10.52202/085713-1454} {\emph {\bibinfo {booktitle} {Advances in Neural Information Processing Systems}}},\ Vol.~\bibinfo {volume} {38}\ (\bibinfo {year} {2025})\ pp.\ \bibinfo {pages} {48750--48801}\BibitemShut {NoStop}%
\bibitem [{\citenamefont {Horsman}\ \emph {et~al.}(2012)\citenamefont {Horsman}, \citenamefont {Fowler}, \citenamefont {Devitt},\ and\ \citenamefont {Van~Meter}}]{horsman2012}%
  \BibitemOpen
  \bibfield  {author} {\bibinfo {author} {\bibfnamefont {D.}~\bibnamefont {Horsman}}, \bibinfo {author} {\bibfnamefont {A.~G.}\ \bibnamefont {Fowler}}, \bibinfo {author} {\bibfnamefont {S.}~\bibnamefont {Devitt}},\ and\ \bibinfo {author} {\bibfnamefont {R.}~\bibnamefont {Van~Meter}},\ }\bibfield  {title} {\bibinfo {title} {Surface code quantum computing by lattice surgery},\ }\href {https://doi.org/10.1088/1367-2630/14/12/123011} {\bibfield  {journal} {\bibinfo  {journal} {New J. Phys.}\ }\textbf {\bibinfo {volume} {14}},\ \bibinfo {pages} {123011} (\bibinfo {year} {2012})}\BibitemShut {NoStop}%
\bibitem [{\citenamefont {Litinski}(2019)}]{litinski2019}%
  \BibitemOpen
  \bibfield  {author} {\bibinfo {author} {\bibfnamefont {D.}~\bibnamefont {Litinski}},\ }\bibfield  {title} {\bibinfo {title} {A game of surface codes: Large-scale quantum computing with lattice surgery},\ }\href {https://doi.org/10.22331/q-2019-03-05-128} {\bibfield  {journal} {\bibinfo  {journal} {Quantum}\ }\textbf {\bibinfo {volume} {3}},\ \bibinfo {pages} {128} (\bibinfo {year} {2019})}\BibitemShut {NoStop}%
\bibitem [{\citenamefont {Erhard}\ \emph {et~al.}(2021)\citenamefont {Erhard} \emph {et~al.}}]{erhard2021}%
  \BibitemOpen
  \bibfield  {author} {\bibinfo {author} {\bibfnamefont {A.}~\bibnamefont {Erhard}} \emph {et~al.},\ }\bibfield  {title} {\bibinfo {title} {Entangling logical qubits with lattice surgery},\ }\href {https://doi.org/10.1038/s41586-020-03079-6} {\bibfield  {journal} {\bibinfo  {journal} {Nature}\ }\textbf {\bibinfo {volume} {589}},\ \bibinfo {pages} {220} (\bibinfo {year} {2021})}\BibitemShut {NoStop}%
\bibitem [{\citenamefont {Besedin}\ \emph {et~al.}(2026)\citenamefont {Besedin} \emph {et~al.}}]{besedin2026}%
  \BibitemOpen
  \bibfield  {author} {\bibinfo {author} {\bibfnamefont {I.}~\bibnamefont {Besedin}} \emph {et~al.},\ }\bibfield  {title} {\bibinfo {title} {Lattice surgery realized on two distance-three repetition codes with superconducting qubits},\ }\href {https://doi.org/10.1038/s41567-025-03090-6} {\bibfield  {journal} {\bibinfo  {journal} {Nat. Phys.}\ }\textbf {\bibinfo {volume} {22}},\ \bibinfo {pages} {189} (\bibinfo {year} {2026})}\BibitemShut {NoStop}%
\bibitem [{\citenamefont {Lin}\ \emph {et~al.}(2026)\citenamefont {Lin} \emph {et~al.}}]{lin2026surgery}%
  \BibitemOpen
  \bibfield  {author} {\bibinfo {author} {\bibfnamefont {W.}~\bibnamefont {Lin}} \emph {et~al.},\ }\href {https://arxiv.org/abs/2607.01473} {\bibinfo {title} {Surface code logical operations on a superconducting quantum processor}} (\bibinfo {year} {2026}),\ \Eprint {https://arxiv.org/abs/2607.01473} {arXiv:2607.01473 [quant-ph]} \BibitemShut {NoStop}%
\bibitem [{\citenamefont {Bausch}\ \emph {et~al.}(2024)\citenamefont {Bausch} \emph {et~al.}}]{bausch2024alphaqubit}%
  \BibitemOpen
  \bibfield  {author} {\bibinfo {author} {\bibfnamefont {J.}~\bibnamefont {Bausch}} \emph {et~al.},\ }\bibfield  {title} {\bibinfo {title} {Learning high-accuracy error decoding for quantum processors},\ }\href {https://doi.org/10.1038/s41586-024-08148-8} {\bibfield  {journal} {\bibinfo  {journal} {Nature}\ }\textbf {\bibinfo {volume} {635}},\ \bibinfo {pages} {834} (\bibinfo {year} {2024})}\BibitemShut {NoStop}%
\bibitem [{\citenamefont {Schreiber}\ \emph {et~al.}(2023)\citenamefont {Schreiber}, \citenamefont {Eisert},\ and\ \citenamefont {Meyer}}]{schreiber2023classical}%
  \BibitemOpen
  \bibfield  {author} {\bibinfo {author} {\bibfnamefont {F.~J.}\ \bibnamefont {Schreiber}}, \bibinfo {author} {\bibfnamefont {J.}~\bibnamefont {Eisert}},\ and\ \bibinfo {author} {\bibfnamefont {J.~J.}\ \bibnamefont {Meyer}},\ }\bibfield  {title} {\bibinfo {title} {Classical surrogates for quantum learning models},\ }\href {https://doi.org/10.1103/PhysRevLett.131.100803} {\bibfield  {journal} {\bibinfo  {journal} {Phys. Rev. Lett.}\ }\textbf {\bibinfo {volume} {131}},\ \bibinfo {pages} {100803} (\bibinfo {year} {2023})}\BibitemShut {NoStop}%
\bibitem [{\citenamefont {Du}\ \emph {et~al.}(2025)\citenamefont {Du}, \citenamefont {Hsieh},\ and\ \citenamefont {Tao}}]{du2025efficient}%
  \BibitemOpen
  \bibfield  {author} {\bibinfo {author} {\bibfnamefont {Y.}~\bibnamefont {Du}}, \bibinfo {author} {\bibfnamefont {M.-H.}\ \bibnamefont {Hsieh}},\ and\ \bibinfo {author} {\bibfnamefont {D.}~\bibnamefont {Tao}},\ }\bibfield  {title} {\bibinfo {title} {Efficient learning for linear properties of bounded-gate quantum circuits},\ }\href {https://doi.org/10.1038/s41467-025-59198-z} {\bibfield  {journal} {\bibinfo  {journal} {Nat. Commun.}\ }\textbf {\bibinfo {volume} {16}},\ \bibinfo {pages} {3790} (\bibinfo {year} {2025})}\BibitemShut {NoStop}%
\bibitem [{\citenamefont {Liao}\ \emph {et~al.}(2026)\citenamefont {Liao}, \citenamefont {Du}, \citenamefont {Wang}, \citenamefont {Tian}, \citenamefont {Luo}, \citenamefont {Du}, \citenamefont {Tao},\ and\ \citenamefont {Huang}}]{liao2026demonstration}%
  \BibitemOpen
  \bibfield  {author} {\bibinfo {author} {\bibfnamefont {W.-Y.}\ \bibnamefont {Liao}}, \bibinfo {author} {\bibfnamefont {Y.}~\bibnamefont {Du}}, \bibinfo {author} {\bibfnamefont {X.}~\bibnamefont {Wang}}, \bibinfo {author} {\bibfnamefont {T.-C.}\ \bibnamefont {Tian}}, \bibinfo {author} {\bibfnamefont {Y.}~\bibnamefont {Luo}}, \bibinfo {author} {\bibfnamefont {B.}~\bibnamefont {Du}}, \bibinfo {author} {\bibfnamefont {D.}~\bibnamefont {Tao}},\ and\ \bibinfo {author} {\bibfnamefont {H.-L.}\ \bibnamefont {Huang}},\ }\bibfield  {title} {\bibinfo {title} {Demonstration of efficient predictive surrogates for large-scale quantum processors},\ }\href {https://doi.org/10.1038/s41467-026-72506-5} {\bibfield  {journal} {\bibinfo  {journal} {Nat. Commun.}\ }\textbf {\bibinfo {volume} {17}},\ \bibinfo {pages} {4731} (\bibinfo {year} {2026})}\BibitemShut {NoStop}%
\bibitem [{\citenamefont {{Harbor Framework Team}}(2026)}]{harbor2026}%
  \BibitemOpen
  \bibfield  {author} {\bibinfo {author} {\bibnamefont {{Harbor Framework Team}}},\ }\href@noop {} {\bibinfo {title} {Harbor: A framework for evaluating and optimizing agents and models in container environments}},\ \bibinfo {howpublished} {\url{https://doi.org/10.5281/zenodo.21878893}} (\bibinfo {year} {2026}),\ \bibinfo {note} {version 0.21.0, \url{https://github.com/harbor-framework/harbor}}\BibitemShut {NoStop}%
\bibitem [{\citenamefont {Merrill}\ \emph {et~al.}(2026)\citenamefont {Merrill}, \citenamefont {Shaw}, \citenamefont {Carlini} \emph {et~al.}}]{merrill2026terminalbench}%
  \BibitemOpen
  \bibfield  {author} {\bibinfo {author} {\bibfnamefont {M.~A.}\ \bibnamefont {Merrill}}, \bibinfo {author} {\bibfnamefont {A.~G.}\ \bibnamefont {Shaw}}, \bibinfo {author} {\bibfnamefont {N.}~\bibnamefont {Carlini}}, \emph {et~al.},\ }\href {https://arxiv.org/abs/2601.11868} {\bibinfo {title} {{Terminal-Bench}: Benchmarking agents on hard, realistic tasks in command line interfaces}} (\bibinfo {year} {2026}),\ \Eprint {https://arxiv.org/abs/2601.11868} {arXiv:2601.11868 [cs.SE]} \BibitemShut {NoStop}%
\bibitem [{\citenamefont {{Terminal-Bench-Science Team}}(2026)}]{terminalbenchscience2026}%
  \BibitemOpen
  \bibfield  {author} {\bibinfo {author} {\bibnamefont {{Terminal-Bench-Science Team}}},\ }\href {https://doi.org/10.5281/zenodo.22110253} {\bibinfo {title} {{Terminal-Bench-Science: Evaluating AI agents on research workflows across scientific domains}}},\ \bibinfo {howpublished} {Zenodo, \url{https://doi.org/10.5281/zenodo.22110253}} (\bibinfo {year} {2026})\BibitemShut {NoStop}%
\bibitem [{\citenamefont {{Anthropic}}(2024)}]{mcp2024}%
  \BibitemOpen
  \bibfield  {author} {\bibinfo {author} {\bibnamefont {{Anthropic}}},\ }\href@noop {} {\bibinfo {title} {Introducing the {Model Context Protocol}}},\ \bibinfo {howpublished} {\url{https://www.anthropic.com/news/model-context-protocol}} (\bibinfo {year} {2024})\BibitemShut {NoStop}%
\bibitem [{\citenamefont {Javadi-Abhari}\ \emph {et~al.}(2024)\citenamefont {Javadi-Abhari}, \citenamefont {Treinish}, \citenamefont {Krsulich}, \citenamefont {Wood}, \citenamefont {Lishman}, \citenamefont {Gacon}, \citenamefont {Martiel}, \citenamefont {Nation}, \citenamefont {Bishop}, \citenamefont {Cross}, \citenamefont {Johnson},\ and\ \citenamefont {Gambetta}}]{javadiabhari2024quantum}%
  \BibitemOpen
  \bibfield  {author} {\bibinfo {author} {\bibfnamefont {A.}~\bibnamefont {Javadi-Abhari}}, \bibinfo {author} {\bibfnamefont {M.}~\bibnamefont {Treinish}}, \bibinfo {author} {\bibfnamefont {K.}~\bibnamefont {Krsulich}}, \bibinfo {author} {\bibfnamefont {C.~J.}\ \bibnamefont {Wood}}, \bibinfo {author} {\bibfnamefont {J.}~\bibnamefont {Lishman}}, \bibinfo {author} {\bibfnamefont {J.}~\bibnamefont {Gacon}}, \bibinfo {author} {\bibfnamefont {S.}~\bibnamefont {Martiel}}, \bibinfo {author} {\bibfnamefont {P.~D.}\ \bibnamefont {Nation}}, \bibinfo {author} {\bibfnamefont {L.~S.}\ \bibnamefont {Bishop}}, \bibinfo {author} {\bibfnamefont {A.~W.}\ \bibnamefont {Cross}}, \bibinfo {author} {\bibfnamefont {B.~R.}\ \bibnamefont {Johnson}},\ and\ \bibinfo {author} {\bibfnamefont {J.~M.}\ \bibnamefont {Gambetta}},\ }\href {https://doi.org/10.48550/arXiv.2405.08810} {\bibinfo {title} {Quantum computing with {Q}iskit}} (\bibinfo {year} {2024}),\ \Eprint {https://arxiv.org/abs/2405.08810} {arXiv:2405.08810 [quant-ph]}
  \BibitemShut {NoStop}%
\bibitem [{\citenamefont {{QCoDeS Contributors}}(2026)}]{qcodes}%
  \BibitemOpen
  \bibfield  {author} {\bibinfo {author} {\bibnamefont {{QCoDeS Contributors}}},\ }\href {https://doi.org/10.5281/zenodo.596989} {\bibinfo {title} {{QCoDeS}: Python-based data acquisition framework}},\ \bibinfo {howpublished} {Zenodo, \url{https://doi.org/10.5281/zenodo.596989}} (\bibinfo {year} {2026})\BibitemShut {NoStop}%
\bibitem [{\citenamefont {Islam}\ \emph {et~al.}(2026)\citenamefont {Islam}, \citenamefont {Wadekar},\ and\ \citenamefont {Zhou}}]{islam2026gwbenchmarks}%
  \BibitemOpen
  \bibfield  {author} {\bibinfo {author} {\bibfnamefont {T.}~\bibnamefont {Islam}}, \bibinfo {author} {\bibfnamefont {D.}~\bibnamefont {Wadekar}},\ and\ \bibinfo {author} {\bibfnamefont {Z.}~\bibnamefont {Zhou}},\ }\href {https://doi.org/10.48550/arXiv.2605.11269} {\bibinfo {title} {{gwBenchmarks}: Stress-testing {LLM} agents on high-precision gravitational wave astronomy}} (\bibinfo {year} {2026}),\ \Eprint {https://arxiv.org/abs/2605.11269} {arXiv:2605.11269 [gr-qc]} \BibitemShut {NoStop}%
\bibitem [{\citenamefont {Chowdhury}\ \emph {et~al.}(2025)\citenamefont {Chowdhury}, \citenamefont {Johnson}, \citenamefont {Huang}, \citenamefont {Steinhardt},\ and\ \citenamefont {Schwettmann}}]{chowdhury2025truthfulness}%
  \BibitemOpen
  \bibfield  {author} {\bibinfo {author} {\bibfnamefont {N.}~\bibnamefont {Chowdhury}}, \bibinfo {author} {\bibfnamefont {D.}~\bibnamefont {Johnson}}, \bibinfo {author} {\bibfnamefont {V.}~\bibnamefont {Huang}}, \bibinfo {author} {\bibfnamefont {J.}~\bibnamefont {Steinhardt}},\ and\ \bibinfo {author} {\bibfnamefont {S.}~\bibnamefont {Schwettmann}},\ }\href@noop {} {\bibinfo {title} {Investigating truthfulness in a pre-release o3 model}},\ \bibinfo {howpublished} {Transluce technical report, \url{https://transluce.org/investigating-o3-truthfulness}} (\bibinfo {year} {2025}),\ \bibinfo {note} {accessed 8 August 2026}\BibitemShut {NoStop}%
\bibitem [{\citenamefont {Jones}\ \emph {et~al.}(2012)\citenamefont {Jones}, \citenamefont {Van~Meter}, \citenamefont {Fowler}, \citenamefont {McMahon}, \citenamefont {Kim}, \citenamefont {Ladd},\ and\ \citenamefont {Yamamoto}}]{Jones2012layered}%
  \BibitemOpen
  \bibfield  {author} {\bibinfo {author} {\bibfnamefont {N.~C.}\ \bibnamefont {Jones}}, \bibinfo {author} {\bibfnamefont {R.}~\bibnamefont {Van~Meter}}, \bibinfo {author} {\bibfnamefont {A.~G.}\ \bibnamefont {Fowler}}, \bibinfo {author} {\bibfnamefont {P.~L.}\ \bibnamefont {McMahon}}, \bibinfo {author} {\bibfnamefont {J.}~\bibnamefont {Kim}}, \bibinfo {author} {\bibfnamefont {T.~D.}\ \bibnamefont {Ladd}},\ and\ \bibinfo {author} {\bibfnamefont {Y.}~\bibnamefont {Yamamoto}},\ }\bibfield  {title} {\bibinfo {title} {Layered architecture for quantum computing},\ }\href {https://doi.org/10.1103/PhysRevX.2.031007} {\bibfield  {journal} {\bibinfo  {journal} {Phys. Rev. X}\ }\textbf {\bibinfo {volume} {2}},\ \bibinfo {pages} {031007} (\bibinfo {year} {2012})}\BibitemShut {NoStop}%
\bibitem [{\citenamefont {Sarovar}\ \emph {et~al.}(2020)\citenamefont {Sarovar}, \citenamefont {Proctor}, \citenamefont {Rudinger}, \citenamefont {Young}, \citenamefont {Nielsen},\ and\ \citenamefont {Blume-Kohout}}]{sarovar2020crosstalk}%
  \BibitemOpen
  \bibfield  {author} {\bibinfo {author} {\bibfnamefont {M.}~\bibnamefont {Sarovar}}, \bibinfo {author} {\bibfnamefont {T.}~\bibnamefont {Proctor}}, \bibinfo {author} {\bibfnamefont {K.}~\bibnamefont {Rudinger}}, \bibinfo {author} {\bibfnamefont {K.}~\bibnamefont {Young}}, \bibinfo {author} {\bibfnamefont {E.}~\bibnamefont {Nielsen}},\ and\ \bibinfo {author} {\bibfnamefont {R.}~\bibnamefont {Blume-Kohout}},\ }\bibfield  {title} {\bibinfo {title} {Detecting crosstalk errors in quantum information processors},\ }\href {https://doi.org/10.22331/q-2020-09-11-321} {\bibfield  {journal} {\bibinfo  {journal} {Quantum}\ }\textbf {\bibinfo {volume} {4}},\ \bibinfo {pages} {321} (\bibinfo {year} {2020})}\BibitemShut {NoStop}%
\bibitem [{\citenamefont {Koch}\ \emph {et~al.}(2022)\citenamefont {Koch}, \citenamefont {Boscain}, \citenamefont {Calarco}, \citenamefont {Dirr}, \citenamefont {Filipp}, \citenamefont {Glaser}, \citenamefont {Kosloff}, \citenamefont {Montangero}, \citenamefont {Schulte-Herbr{\"u}ggen}, \citenamefont {Sugny},\ and\ \citenamefont {Wilhelm}}]{koch2022quantumcontrol}%
  \BibitemOpen
  \bibfield  {author} {\bibinfo {author} {\bibfnamefont {C.~P.}\ \bibnamefont {Koch}}, \bibinfo {author} {\bibfnamefont {U.}~\bibnamefont {Boscain}}, \bibinfo {author} {\bibfnamefont {T.}~\bibnamefont {Calarco}}, \bibinfo {author} {\bibfnamefont {G.}~\bibnamefont {Dirr}}, \bibinfo {author} {\bibfnamefont {S.}~\bibnamefont {Filipp}}, \bibinfo {author} {\bibfnamefont {S.~J.}\ \bibnamefont {Glaser}}, \bibinfo {author} {\bibfnamefont {R.}~\bibnamefont {Kosloff}}, \bibinfo {author} {\bibfnamefont {S.}~\bibnamefont {Montangero}}, \bibinfo {author} {\bibfnamefont {T.}~\bibnamefont {Schulte-Herbr{\"u}ggen}}, \bibinfo {author} {\bibfnamefont {D.}~\bibnamefont {Sugny}},\ and\ \bibinfo {author} {\bibfnamefont {F.~K.}\ \bibnamefont {Wilhelm}},\ }\bibfield  {title} {\bibinfo {title} {Quantum optimal control in quantum technologies. strategic report on current status, visions and goals for research in europe},\ }\href {https://doi.org/10.1140/epjqt/s40507-022-00138-x} {\bibfield  {journal} {\bibinfo  {journal} {EPJ Quantum
  Technol.}\ }\textbf {\bibinfo {volume} {9}},\ \bibinfo {pages} {19} (\bibinfo {year} {2022})}\BibitemShut {NoStop}%
\bibitem [{\citenamefont {de~Vega}\ and\ \citenamefont {Alonso}(2017)}]{devega2017dynamics}%
  \BibitemOpen
  \bibfield  {author} {\bibinfo {author} {\bibfnamefont {I.}~\bibnamefont {de~Vega}}\ and\ \bibinfo {author} {\bibfnamefont {D.}~\bibnamefont {Alonso}},\ }\bibfield  {title} {\bibinfo {title} {Dynamics of non-markovian open quantum systems},\ }\href {https://doi.org/10.1103/RevModPhys.89.015001} {\bibfield  {journal} {\bibinfo  {journal} {Rev. Mod. Phys.}\ }\textbf {\bibinfo {volume} {89}},\ \bibinfo {pages} {015001} (\bibinfo {year} {2017})}\BibitemShut {NoStop}%
\bibitem [{\citenamefont {Briegel}\ \emph {et~al.}(1998)\citenamefont {Briegel}, \citenamefont {D\"ur}, \citenamefont {Cirac},\ and\ \citenamefont {Zoller}}]{briegel1998repeaters}%
  \BibitemOpen
  \bibfield  {author} {\bibinfo {author} {\bibfnamefont {H.-J.}\ \bibnamefont {Briegel}}, \bibinfo {author} {\bibfnamefont {W.}~\bibnamefont {D\"ur}}, \bibinfo {author} {\bibfnamefont {J.~I.}\ \bibnamefont {Cirac}},\ and\ \bibinfo {author} {\bibfnamefont {P.}~\bibnamefont {Zoller}},\ }\bibfield  {title} {\bibinfo {title} {Quantum repeaters: The role of imperfect local operations in quantum communication},\ }\href {https://doi.org/10.1103/PhysRevLett.81.5932} {\bibfield  {journal} {\bibinfo  {journal} {Phys. Rev. Lett.}\ }\textbf {\bibinfo {volume} {81}},\ \bibinfo {pages} {5932} (\bibinfo {year} {1998})}\BibitemShut {NoStop}%
\bibitem [{\citenamefont {Chin}\ \emph {et~al.}(2012)\citenamefont {Chin}, \citenamefont {Huelga},\ and\ \citenamefont {Plenio}}]{chin2012metrology}%
  \BibitemOpen
  \bibfield  {author} {\bibinfo {author} {\bibfnamefont {A.~W.}\ \bibnamefont {Chin}}, \bibinfo {author} {\bibfnamefont {S.~F.}\ \bibnamefont {Huelga}},\ and\ \bibinfo {author} {\bibfnamefont {M.~B.}\ \bibnamefont {Plenio}},\ }\bibfield  {title} {\bibinfo {title} {Quantum metrology in non-markovian environments},\ }\href {https://doi.org/10.1103/PhysRevLett.109.233601} {\bibfield  {journal} {\bibinfo  {journal} {Phys. Rev. Lett.}\ }\textbf {\bibinfo {volume} {109}},\ \bibinfo {pages} {233601} (\bibinfo {year} {2012})}\BibitemShut {NoStop}%
\bibitem [{\citenamefont {Shepard}\ and\ \citenamefont {Salimans}(2026)}]{zapier2026automationbench}%
  \BibitemOpen
  \bibfield  {author} {\bibinfo {author} {\bibfnamefont {D.}~\bibnamefont {Shepard}}\ and\ \bibinfo {author} {\bibfnamefont {R.}~\bibnamefont {Salimans}},\ }\href@noop {} {\bibinfo {title} {{AutomationBench}}} (\bibinfo {year} {2026}),\ \bibinfo {note} {benchmark by Zapier; scores quoted in this work are from the Artificial Analysis independent run, not from this paper, whose strict whole-task completion metric reports different values},\ \Eprint {https://arxiv.org/abs/2604.18934} {arXiv:2604.18934} \BibitemShut {NoStop}%
\bibitem [{\citenamefont {{Artificial Analysis}}(2026{\natexlab{a}})}]{artificialanalysis2026automationbench}%
  \BibitemOpen
  \bibfield  {author} {\bibinfo {author} {\bibnamefont {{Artificial Analysis}}},\ }\href@noop {} {\bibinfo {title} {{AutomationBench-AA}: Independent evaluation of {AutomationBench}}},\ \bibinfo {howpublished} {\url{https://artificialanalysis.ai/evaluations/automationbench-aa}} (\bibinfo {year} {2026}{\natexlab{a}}),\ \bibinfo {note} {independent run by Artificial Analysis on a private held-out set of 657 tasks across six business domains; the headline metric is the average share of each task's objectives completed without guardrail violations. Accessed 2026-09-15.}\BibitemShut {Stop}%
\bibitem [{\citenamefont {{Artificial Analysis}}(2026{\natexlab{b}})}]{artificialanalysis2026gdpval}%
  \BibitemOpen
  \bibfield  {author} {\bibinfo {author} {\bibnamefont {{Artificial Analysis}}},\ }\href@noop {} {\bibinfo {title} {{GDPval-AA v2}: Independent evaluation of {OpenAI}'s {GDPval} gold set}},\ \bibinfo {howpublished} {\url{https://artificialanalysis.ai/evaluations/gdpval-aa}} (\bibinfo {year} {2026}{\natexlab{b}}),\ \bibinfo {note} {{Elo} ratings from blind pairwise comparisons; agents run with shell and browsing access via Stirrup. Accessed 2026-09-15.}\BibitemShut {Stop}%
\end{thebibliography}

%

\end{document}